\documentclass[pdftex,twocolumn,epjc3]{svjour3}          

\smartqed  
\RequirePackage{graphicx}

\newcommand{\be}{\begin{equation}}
\newcommand{\ee}{\end{equation}}
\newcommand{\ba}{\begin{Eqnarray}}
\newcommand{\ea}{\end{Eqnarray}}

\def\makeheadbox{{\hfill  CFTP/20-003}}
\def\rubric{{}}

\usepackage{amsmath,amsbsy,amsfonts,amsmath,bbm,latexsym,amssymb,bm}

\begin{document}

\title{Nondecoupling in Multi-Higgs doublet models}


\author{Francisco Faro\thanksref{e1,addr1} \and
   Jorge C. Rom\~{a}o\thanksref{e2,addr1} \and
    Jo\~{a}o P. Silva\thanksref{e3,addr1}}

\thankstext{e1}{e-mail: francisco.faro@tecnico.ulisboa.pt}
\thankstext{e2}{e-mail: jorge.romao@tecnico.ulisboa.pt}
\thankstext{e3}{e-mail: jpsilva@cftp.ist.utl.pt}
\institute{Departamento de F\'{\i}sica and Centro de F\'{\i}sica Te\'{o}rica de
Part\'{\i}culas (CFTP),\\
\quad Instituto Superior T\'{e}cnico (IST), U. de Lisboa (UL),\\ 
\quad Av. Rovisco Pais, P-1049-001 Lisboa, Portugal.\label{addr1}}

\date{\today}

\maketitle

\begin{abstract}
We consider models with any number of Higgs doublets and study the
conditions for decoupling. We show that, under very general circumstances,
all the quadratic coefficients of the scalar potential must be present,
except in special cases, which include terms related to directions of
vanishing vacuum expectation values. We give a few examples.
Moreover, we show that the decoupling of all charged scalars implies the
decoupling of all extra neutral scalars and vanishing $\mathcal{CP}$ violation
in scalar-pseudoscalar mixing.
\end{abstract}


\section{Introduction}
\label{sec:intro}

It is has been determined experimentally that there are four electroweak gauge
bosons \cite{Arnison:1983rp,Arnison:1983mk,Bagnaia:1983zx,Banner:1983jy},
as predicted by the $\mathbb{SU}(2)_L \times \mathbb{U}(1)_Y$
gauge group of the Standard Model (SM).
The SM gauge structure does not predict the number of fermion families,
which was established at LEP and SLD by the invisible width
of the $Z$ \cite{ALEPH:2005ab}.
Neither does the SM structure predict the number of scalar doublet families.
This is the most fundamental open question,
and it is being actively pursued at LHC.

The Atlas and CMS experiments have already identified
a $125\, \textrm{GeV}$ Higgs particle ($h_{125}$)
\cite{Aad:2012tfa,Chatrchyan:2012xdj},
as established in the (minimal $N=1$ Higgs version of the) SM,
and have checked that its couplings are consistent
with those predicted, within errors of order 20\% \cite{Khachatryan:2016vau}.
In N Higgs doublet models (NHDM) there are in general
$2N-1$ neutral scalars $S_\alpha^0$ ($\alpha = 2, \dots 2N$)
and $N-1$ charged scalar pairs $S^\pm_a$ ($a = 2, \dots N$).\footnote{We
keep the first entries $S_1^0=G^0$ and $S_1^\pm=G^\pm$ for the would-be
Goldstone bosons.}
States beyond the $S^0_2=h_{125}$ scalar have not been found,
nor its number limited.
The reason is that the generic NHDM has a decoupling limit,
where the extra scalars have very high masses and the remaining
$h_{125}$ has basically the properties of the SM one.

There are, however, NHDMs in which the extra Higgses
do not decouple. That is the case, for example, with the
2HDM with an exact $Z_2$ symmetry \cite{Glashow:1976nt}.
Nevertheless,
Gunion and Haber \cite{Gunion:2002zf} have shown that the decoupling limit
is recovered by including in the potential a term of dimension two
which breaks softly the $Z_2$ symmetry.
Nondecoupling has  also been analyzed by Nebot in the 2HDM with
spontaneous $\mathcal{CP}$ violation, both with and without soft symmetry
breaking terms \cite{Nebot:2019lzf},
with interesting physical consequences explored by
Nierste, Tabet, and Ziegler \cite{Nierste:2019fbx}.
One further example with 3HDM and a $S_3$ symmetry has been
discussed by Bhattacharyya and Das \cite{Bhattacharyya:2014oka}.
In this article we discuss the situation in the NHDM.
If these have an exact symmetry limiting the number of quadratic and
quartic couplings, then one can expect that nondecoupling occurs.
Conversely, as we show in this article,
for the existence of a decoupling limit,
all soft breaking terms must be included in the scalar potential,
except when there are small mixing angles in the scalar sector,
which includes directions with vanishing vacuum expectation
values.
We introduce our notation in section~\ref{sec:notation}, 
where we also present our first results.
Section~\ref{sec:decoupling} is devoted to theorems valid when
there are no vanishing vacuum expectation values (vev)
and/or small mixing angles.
We present a few 3HDM examples in section~\ref{sec:examples},
both of the theorems and of what happens when
the assumptions of the theorems are violated.
We discuss briefly $\mathcal{CP}$ violation in section~\ref{sec:cpv}.
In section~\ref{sec:hypothesis} we derive one further
theorem, valid when only one vev is nonzero,
and we present a complete  discussion of all
symmetric 2HDM.
We draw our conclusions in section~\ref{sec:conclusions}.
Some detailed discussions have been relegated to the
Appendices.

\section{\label{sec:notation}Notation and first results}

Consider a  $\mathbb{SU}(2)_L \times \mathbb{U}(1)_Y$ theory with
$N$ complex scalar doublets $\Phi_i$ with hypercharge $Y=1/2$.

\subsection{The scalar potential}

Following the notation in
\cite{Botella:1994cs,Barroso:2006pa,Bento:2017eti},
we write the scalar potential as
\begin{equation}
V_H =
Y_{ij}\left(\Phi_i^\dagger \Phi_j\right)
+ Z_{ij,kl} \left(\Phi_i^\dagger \Phi_j\right)\left(\Phi_k^\dagger \Phi_l\right),
\label{eq:Nhdm_potential}
\end{equation}
whose hermiticity implies,
\begin{equation}
    Y_{ij}=Y_{ji}^* , \quad
    Z_{ij,kl}=Z_{kl,ij}=Z_{ji,lk}^*.
\end{equation}
Requiring a massless photon implies that the
global minimum corresponds to vacuum expectation values (vev)
which preserve the charge symmetry, $\mathbb{U}(1)_Q$, generated by $Q=T_3+Y$.
Expanding the fields around those vevs $\nu_i$,
we write
\begin{equation}
\Phi_i = \nu_i + \varphi_i =
\begin{pmatrix}
0 \\ v_i/\sqrt{2}
\end{pmatrix}
+
\begin{pmatrix}
\phi^+_i \\ \left(\rho_i+i\chi_i\right)/\sqrt{2}
\end{pmatrix},
\label{eq:expansion_fields}
\end{equation}
where each $v_i$ is in general complex.
The stationary conditions are given by
\begin{equation}
\left(Y_{ij}+Z_{ij,kl}v_k^* v_l\right)v_j=0.
\label{eq:Nhdm_minimisation_conditions}
\end{equation}
Implicit summation of repeated indices is used throughout,
except where noted otherwise.
The mass matrix for the charged scalars is,
\begin{equation}
\left(M^2_\pm\right)_{ij} =
Y_{ij}+Z_{ij,kl}v^{*}_k v_l,
\label{eq:Nhdm_mass_matrix_charged}
\end{equation}
in terms of which the stationarity conditions in
\eqref{eq:Nhdm_minimisation_conditions}
may be written as
\be
\left(M^2_\pm\right)_{ij} v_j = 0.
\label{eq:null_mass_state}
\ee
The mass matrix for the neutral scalars is,
\begin{align}
M_{0}^2=&
\begin{pmatrix}
M^2_{\rho\rho} & M^2_{\rho\chi} \\
M^2_{\chi\rho} & M^2_{\chi\chi}
\end{pmatrix},
\label{eq:Nhdm_mass_matrix_neutral}\\
\left(M^2_{\rho\rho}\right)_{ij}=&\,
\text{Re}\left\{ Y_{ij} + Z_{ij,kl}v^{*}_k v_l \right.
\nonumber\\
 & \hspace{5mm}
\left.
+\, Z_{ik,lj}v_k v^{*}_l + Z_{ik,jl}v_k v_l\right\} ,
\label{eq:Nhdm_mass_matrix_R} \\
\left(M^2_{\chi\chi}\right)_{ij}=&\,
\text{Re}\left\{ Y_{ij} + Z_{ij,kl}v^{*}_k v_l \right.
\nonumber\\
 & \hspace{5mm}
\left.
+\, 
Z_{ik,lj}v_k v^{*}_l - Z_{ik,jl}v_k v_l\right\} ,
\label{eq:Nhdm_mass_matrix_I} \\
-\left(M^2_{\rho\chi}\right)_{ij}=&\,
\text{Im}\left\{ Y_{ij} + Z_{ij,kl}v^{*}_k v_l\right.
\nonumber\\
 & \hspace{5mm}
\left.
+\, 
Z_{ik,lj}v_k v^{*}_l - Z_{ik,jl}v_k v_l\right\},
\label{eq:Nhdm_mass_matrix_RI} \\
\left(M^2_{\chi\rho}\right)_{ij}=&\,
\text{Im}\left\{ Y_{ij} + Z_{ij,kl}v^{*}_k v_l \right.
\nonumber\\
 & \hspace{5mm}
\left.
+\, 
Z_{ik,lj}v_k v^{*}_l + Z_{ik,jl}v_k v_l\right\} ,
\label{eq:Nhdm_mass_matrix_IR}
\end{align}
where,
under the canonical definition of $\mathcal{CP}$,
$M^2_{\rho\rho}$ is the mass matrix of the
$\mathcal{CP}$-even scalars,
$M^2_{\chi\chi}$ of the $\mathcal{CP}$-odd scalars,
and $M^2_{\rho\chi}=\left(M^2_{\chi\rho}\right)^T$
gives the mixing between the $\mathcal{CP}$-even and $\mathcal{CP}$-odd scalars.
Substituting \eqref{eq:Nhdm_mass_matrix_charged},
we obtain
\begin{align}
\left(M^2_{\rho\rho}\right)_{ij}=&\,
\text{Re}\left\{ \left(M^2_\pm\right)_{ij}
+ Z_{ik,lj}v_k v^{*}_l + Z_{ik,jl}v_k v_l\right\} ,
\label{eq:Nhdm_mass_matrix_R_v2} \\
\left( M^2_{\chi\chi}\right)_{ij}=&\,
\text{Re}\left\{ \left(M^2_\pm\right)_{ij}
+
Z_{ik,lj}v_k v^{*}_l - Z_{ik,jl}v_k v_l\right\} ,
\label{eq:Nhdm_mass_matrix_I_v2} \\
-\left(M^2_{\rho\chi}\right)_{ij}=&\,
\text{Im}\left\{\left(M^2_\pm\right)_{ij}
+
Z_{ik,lj}v_k v^{*}_l - Z_{ik,jl}v_k v_l\right\} .
\label{eq:Nhdm_mass_matrix_RI_v2}
\end{align}

\subsection{\label{subsec:basis}Basis freedom and symmetries}

We may choose to describe the theory in terms of new fields
$\Phi^\prime_i$, obtained through a basis transformation
which leaves the kinetic terms unchanged
\begin{equation}
\Phi_i \rightarrow \Phi_i^{\prime}= U_{ij} \Phi_j,
\label{basis_transformation}
\end{equation}
where $U$ is an $N \times N$ unitary matrix.
Since the theory is invariant under a global $\mathbb{U}(1)$,
we may take $U$ in $\mathbb{SU}(N)$.
Under this basis change,
the potential parameters and the vevs are transformed as,
\begin{align}
Y_{ij}&
\rightarrow Y_{ij}^{\prime}= U_{ik}Y_{kl}U_{jl}^*,
\label{eq:Nhdm_basis_change_quadratic} \\
Z_{ij,kl}&
\rightarrow Z_{ij,kl}^{\prime}=U_{im}U_{ko}Z_{mn,op}U_{jn}^*U_{lp}^*,
\label{eq:Nhdm_basis_change_quartic} \\
v_i& \rightarrow v^{\prime}_i= U_{ij} v_j.
\end{align}
This means that not all parameters have physical significance;
only basis invariant combinations can be observed experimentally
\cite{Botella:1994cs}.

Still, any model with more than one scalar has many free parameters.
Most often, these are curtailed by invoking some specific symmetry
\begin{equation}
\Phi_i \rightarrow \Phi_i^S= S_{ij} \Phi_j,
\label{S_symmetry}
\end{equation}
also given by a $\mathbb{U}(N)$ matrix,
which imposes relations among the potential parameters,
\begin{align}
Y_{ij}&
=Y_{ij}^{S}= S_{ik}Y_{kl}S_{jl}^*,
\label{eq:Nhdm_symmetry_quadratic}\\
Z_{ij,kl}&
=Z_{ij,kl}^{S}=S_{im}S_{ko}Z_{mn,op}S_{jn}^*S_{lp}^*.
\label{eq:Nhdm_symmetry_quartic}
\end{align}
Recall that in a basis change the potential parameters do not
remain the same, whereas under a symmetry these must remain
invariant.\footnote{In both situations the scalar potential is unaffected.}
The symmetry may (or not) be spontaneously broken,
according to whether (or not)
\begin{equation}
v_i = v_i^S= S_{ij} v_j.
\end{equation}

Consider a theory in which $V_H$, when written in terms of the fields
$\Phi_i$, has the symmetry $S$.
Now, perform the basis transformation in eq.~\eqref{basis_transformation}.
When written in terms of the new fields $\Phi^\prime_i$,
$V_H$ is no longer invariant under $S$; rather, it is now invariant
under
\begin{equation}
S^\prime=U S U^\dagger.
\label{Sprime}
\end{equation}
As an example,
the $\mathbb{Z}_2$ 2HDM symmetry
\begin{equation}
\Phi_1 \rightarrow \Phi_1,\ \ \ 
\Phi_2 \rightarrow -\Phi_2,\ \ \
\end{equation}
becomes
\begin{equation}
\Phi^\prime_1 \leftrightarrow \Phi^\prime_2,\ \ \ 
\end{equation}
in the basis
\begin{equation}
\Phi^\prime_1 = \frac{1}{\sqrt{2}} \left( \Phi_1 + \Phi_2\right),
\ \ \ 
\Phi^\prime_2 = \frac{1}{\sqrt{2}} \left( \Phi_1 - \Phi_2\right).
\end{equation}

Eq.~\eqref{Sprime} is a conjugacy relation within $\mathbb{U}(N)$.
In a suitable basis,
$S$ may be brought to the form \cite{Ferreira:2008zy}
\begin{equation}
\left(
\begin{array}{cccc}
e^{i \theta_1} & & & \\
 & e^{i \theta_2}& & \\
 & & \ddots & \\
& & & e^{i \theta_N} 
\end{array}
\right),
\label{S-diagonal-1}
\end{equation}
or, using the invariance under global hypercharge,
\begin{equation}
\left(
\begin{array}{cccc}
\: 1 \: & & & \\
 & e^{i \theta_2}& & \\
 & & \ddots & \\
& & & e^{i \theta_N} 
\end{array}
\right).
\label{S-diagonal-2}
\end{equation}
This choice makes the presence of the symmetry in the Higgs
potential more transparent.

Strictly speaking,
we have just described the situation where there is only one
generator $S$.\footnote{We are using here ``generator'' in the assertion
often used in connection with the ``presentation of a discrete group'' and
not in connection with the algebra of a continuous Lie group.}
Imagine that there are two generators $S_1$ and $S_2$.
Then, it may be that one can bring both generators to diagonal form.
In that case, the symmetry generated by both, ${\cal S}=\{S_1, S_2\}$, is Abelian
and easy to guess from the form of $V_H$ when it is written
in the basis where both generators are diagonal.
If $S_1$ and $S_2$ do not commute, then
${\cal S}=\{S_1, S_2\}$ is a non-Abelian subset of
$\mathbb{U}(N)$.\footnote{We leave a subtlety for \ref{app:subtlety}.}
One can diagonalize $S_1$, or $S_2$, but not both.

\subsection{\label{subsec:SB_CHB}The symmetry basis and the charged Higgs basis}

The basis freedom in eq.~\eqref{basis_transformation}
may be used to study
a given theory in a specific basis.
There is always a specially simple basis,
the so-called \textit{Charged Higgs basis} (CH basis),
in which the $\mathbb{U}(N)$ basis freedom is used in order
to diagonalize the mass matrix of the charged Higgs in
eq.~\eqref{eq:Nhdm_mass_matrix_charged}
\cite{Bento:2017eti,Nishi:2007nh}.
In this basis,
$v_1=v$, $v_{i\neq 1}=0$,
and the fields may be parametrized as
\begin{align}
\Phi^\textrm{CH}_1 &=
\left(
\begin{array}{c}
G^+\\*[2mm]
\tfrac{1}{\sqrt{2}}
\left( v + H^0 + i G^0 \right)
\end{array}
\right),
\nonumber\\
\Phi^\textrm{CH}_2 &=
\left(
\begin{array}{c}
S^+_2\\*[2mm]
\tfrac{1}{\sqrt{2}} \varphi^{C0}_2
\end{array}
\right)\,,
\nonumber\\
\vdots &
\nonumber\\
\Phi^\textrm{CH}_N &=
\left(
\begin{array}{c}
S^+_N\\*[2mm]
\tfrac{1}{\sqrt{2}} \varphi^{C0}_N
\end{array}
\right),
\label{chhiggsbasis}
\end{align}
where $S_2^+, \ldots, S^+_N$ are the physical charged Higgs mass eigenstate fields,
with corresponding masses $m^2_{\pm, i}$,
and 
\begin{equation}
Y_{ij}^\textrm{CH}+Z_{ij,11}^\textrm{CH} v^2 = \delta_{ij}\, m_{\pm, i}^2
\ \ \ \textrm{(no sum)}.
\label{eq:Nhdm_mass_matrix_charged_CH}
\end{equation}
$S_1^\pm=G^\pm$ is the would-be Goldstone boson with $m_{\pm, 1}^2=0$.
This basis is especially adapted to study the decoupling limit.
In this limit,
all (massive) charged Higgs ($S^\pm_{i \neq 1}$) should acquire 
a very large mass. 
Since perturbativity and unitarity constrains the quartic parameters to lie below
some upper bound, which one may take to be of order $4 \pi \sim O(10)$,
we find
\begin{align}
Y_{11}^\textrm{CH} + Z_{11,11}^\textrm{CH} v^2 &
= 0
& \Rightarrow  &\ \ Y_{11}^\textrm{CH} \sim v^2,
\\
Y_{i\ j\neq i}^\textrm{CH} + Z_{ij,11}^\textrm{CH} v^2 &
= 0
& \Rightarrow  &\ \ Y_{i\ j\neq i}^\textrm{CH} \sim v^2,
\\
Y_{j\ j\neq 1}^\textrm{CH} + Z_{jj,11}^\textrm{CH} v^2 &
= m_{\pm, j}^2
& \Rightarrow   &\ \ Y_{j\ j\neq 1}^\textrm{CH} \sim  m_{\pm, j\neq 1}^2 \gg v^2.
\label{aux3}
\end{align}
Substituting
eq.~\eqref{eq:Nhdm_mass_matrix_charged_CH} in
eqs.~\eqref{eq:Nhdm_mass_matrix_R_v2}-\eqref{eq:Nhdm_mass_matrix_RI_v2},
one obtains \cite{Bento:2017eti}
\begin{align}
(M_{\rho\rho}^2)_{ij} &=
\delta_{ij}\, m^2_{\pm, i} + v^2\,
\textrm{Re}\left\{ Z_{i1,1j} + Z_{i1,j1} \right\},
\label{MR_CH}\\
\left(M^2_{\chi\chi}\right)_{ij} &=
\delta_{ij}\, m^2_{\pm, i} + v^2\,
\textrm{Re}\left\{ Z_{i1,1j} - Z_{i1,j1} \right\},
\label{MI_CH}\\
\left(M^2_{\rho\chi}\right)_{ij} &=
- v^2\, \textrm{Im}\left\{ Z_{i1,1j} -Z_{i1,j1} \right\},
\label{MRI_CH}
\end{align}
were no sum is implied.
This leads to our first important results.
First,
as the charged Higgs masses become very large,
all neutral particles also become very massive.
This is easy to understand.
As $Y_{i\ i\neq 1}^\textrm{CH}$ becomes positive and very large,
\begin{equation}
{\cal M}_i^2 \equiv Y_{i\ i\neq 1}^\textrm{CH} \gg v^2,
\label{eq:calMi}
\end{equation}
the whole doublet $\Phi^\textrm{CH}_i$ decouples from the rest.
Thus, within the framework of effective field theory,
${\cal M}_i^2$ is the interesting measure of decoupling -- see, for example
\cite{Belusca-Maito:2016dqe}.
Second,
if all charged Higgs become very massive,
then 
$M_{\rho\rho}^2 $ and $M_{\chi\chi}^2 $ become very large and almost diagonal,
while $M_{\rho\chi}^2 $ remains of (small) order $v^2$.
Thus, all  $\mathcal{CP}$ violation in scalar-pseudoscalar mixing vanishes.
This could be viewed as our \textbf{Theorem 0}.

At first sight, this second result may seem puzzling.
Consider two very heavy Higgs doublets.
Can't there be scalar-pseudoscalar mixing among those two heavy doublets?
No!
To understand the reason, we notice that eqs.~\eqref{MR_CH}-\eqref{MRI_CH}
mean that, in order to have significant  $\mathcal{CP}$
violation in scalar-pseudoscalar mixing,
the charged Higgs masses $m_{\pm, i\neq 1}^2$ must be of order
$v^2 = |v_1|^2 \dots |v_N|^2$.
Now, the vev of each doublet is bounded by $|v_j| \leq v$ and,
thus, it cannot be very large.
With vevs of order $v$, one is left with two options.
Charged Higgs masses of order $v$ and possible large scalar-pseudoscalar mixing,
or alternatively, charged Higgs masses much larger than $v$
and necessarily small scalar-pseudoscalar mixing.
Ref.~\cite{Bento:2017eti} showed how crucial the CH basis was in
interpreting unitarity bounds.
The two results presented above provide another striking example
of the usefulness of this basis.
Indeed,
these general results would be very difficult to guess in a generic basis.

For the most general NHDM,
the $Y^\textrm{CH}$ and $Z^\textrm{CH}$ parameters are
free to take any value
(consistent with perturbativity and unitarity).
However,
such models suffer from several problems.
On the one hand, they have too many free parameters and their
study (bounds from current experiments and proposed new signals)
is effectively very difficult.
On the other hand,
such models tend to lead to very large scalar flavour changing
neutral couplings (sFCNC) with fermions,
which are very tightly bound by flavour experiments.
There are three solutions:
make the new scalar masses large (precisely the decoupling limit);
take the sFCNC small;
or, make the sFCNC exactly zero by enforcing some symmetry.
We now focus on models with some symmetry $S$.
As seen in
eqs.~\eqref{eq:Nhdm_symmetry_quadratic}-\eqref{eq:Nhdm_symmetry_quartic},
such a symmetry imposes conditions on the parameters
of $V_H$.
Eq.~\eqref{Sprime} implies that the specific
constraint depends on which representation is chosen for the symmetry.
We consider some specific representation for the symmetry and
name the basis where the symmetry has that particular form as
the \textit{Symmetry basis} (S basis).
And, because it constrains the available parameter space,
such a symmetry will have an impact on the nature of the
parameters in the CH basis.

The CH equations \eqref{eq:Nhdm_mass_matrix_charged_CH}-\eqref{aux3} and \eqref{eq:calMi}
can be written compactly as
\begin{equation}
Y^{CH}_{ij}
	=\mathcal{M}^2_{i}(1-\delta_{i1}\delta_{j1})\delta_{ij} + \Omega^{CH}_{ij}, 
	\ \ \  \textrm{(no sum)}
\label{eq:Nhdm_CHiggsb_quadratic}
\end{equation}
where
\begin{equation}
\Omega^{CH}_{ij}
	=-Z^{CH}_{ij,11}v^2\left(1-\delta_{ij}\right)
	- Z^{CH}_{11,11}v^2\delta_{i1}\delta_{j1}. \ \ \ \textrm{(no sum)}
	\label{eq:Nhdm_CHiggsb_stationary}
\end{equation}
Recall that as $\mathcal{M}^2_{i}$ increases,
it drives the decoupling and defines the energy scale
of the states within the $\Phi^{CH}_{k\neq 1}$ doublet.
The CH basis and the S basis are related by
\begin{align}
\Phi^{CH}
	&=U^{CH}\Phi^S,
\label{eq:Nhdm_CHiggsb_basis_change_1}\\
U^{CH}_{1j}
	&=\frac{v^*_j}{v}, 
\label{eq:Nhdm_CHiggsb_basis_change_2}\\*[2mm]
Y^{S}_{ij}
	&=U^{CH*}_{ki}Y^{CH}_{kl}U^{CH}_{lj},
\label{eq:Nhdm_CHiggsb_basis_change_quadratic} \\*[2mm]
Z^{CH}_{ij,kl}
	&=U^{CH}_{im}U^{CH}_{ko} Z^S_{mn,op} U^{CH*}_{jn} U^{CH*}_{lp}.
\label{eq:Nhdm_CHiggsb_quartic}
\end{align}
Eqs.~\eqref{eq:Nhdm_CHiggsb_quadratic} and \eqref{eq:Nhdm_CHiggsb_basis_change_quadratic}
can be combined into
\begin{equation}
Y^S_{ij}= \sum^N_{k=2} U^{CH*}_{ki} \mathcal{M}^2_{k} U^{CH}_{kj}
+ \sum^N_{k=1,l=1} U^{CH*}_{ki} \Omega^{CH}_{kl} U^{CH}_{lj}.
\label{eq:Nhdm_Sb_quadratic}
\end{equation}
This equation shows explicitly how the quadratic parameters in the S basis
depend on the decoupling parameters $\mathcal{M}^2_{k \neq 1}$.
In order to have a decoupling limit with all $\mathcal{M}^2_{k \neq 1} \gg v^2$
it is certainly sufficient to include all  $Y^S_{ij}$.
However, most symmetries preclude some of these quadratic coefficients,
possibly precluding a decoupling limit.
Our aim is to study when this can and when it cannot happen.

Recall that the matrix $U^{CH}$ is the unitary transformation which
diagonalizes the mass matrix of the charged scalars in
eq.~\eqref{eq:Nhdm_mass_matrix_charged},
when written in the $S$ basis.
The first line of $U^{CH}$ must have the form in
eq.~\eqref{eq:Nhdm_CHiggsb_basis_change_2} because,
through eq.~\eqref{eq:null_mass_state},
that guarantees that the first eigenvector has zero mass,
corresponding to $G^+$.
Eq.~\eqref{eq:Nhdm_CHiggsb_basis_change_2} also guarantees that the
vev of the first doublet in the CH basis is $v$, while all other
doublets in the CH basis have zero vev.
A generic $N \times N$ unitary matrix is defined by 
the $N(N-1)/2$ angles in an orthogonal  $N \times N$  matrix,
and by $N(N+1)/2$ phases.
From the original $N(N-1)/2$ angles, $(N-1)$ are determined
by the vevs, which appear in the first line of $U^{CH}$,
\textit{c.f.} eq.~\eqref{eq:Nhdm_CHiggsb_basis_change_2}.
We denote such angles by $\beta_i$ and, in our notation,
they cannot be multiples of $\pi/2$ if we wish to keep all vevs
different from zero.
There remain $(N-1)(N-2)/2$ angles,
which we denote by $\omega_i$.
In contrast with the $\beta_i$,
these $\omega_i$ angles depend not only on the vevs,
but also on the independent $Y^{S}_{ij}$ and $Z^S_{mn,op}$.
We may choose regions of parameter space such that some
$\omega_i$, and, thus, some entries in the $U^{CH}$ matrix are zero. 

Consider a real 3HDM, with real vevs.
Then,
the unitary transformation from the
S basis to the CH basis can be written as,
\begin{align}
U^{CH} &= \frac{1}{v v_{12}}\begin{pmatrix}
 1 & 0 & 0 \\
 0 & \cos{(\omega)}  & \sin{(\omega)} \\
 0 & -\sin{(\omega)} & \cos{(\omega)} \end{pmatrix}
\nonumber\\
&\hspace{8mm} \times \begin{pmatrix}
    v_{12} & 0       & v_3\\ 
    0      & v & 0\\
    -v_3   & 0       & v_{12}\\
    \end{pmatrix}\begin{pmatrix}
    v_1 & v_2 & 0 \\ 
    -v_2  & v_1   & 0 \\
    0     & 0     & v_{12}\\
    \end{pmatrix}.
\end{align}
Parametrizing the vevs as
\begin{align}
    v_1 &=v\sin(\beta_2)\cos(\beta_1),
\nonumber\\
	v_2 &=v\sin(\beta_2)\sin(\beta_1),
\nonumber\\
	v_3 &=v\cos(\beta_2),
\nonumber\\
	v_{12} &=\sqrt{|v_1|^2 + |v_2|^2}=v\sin(\beta_2),
\label{vevs_123}
\end{align}
one obtains
\begin{equation}
U^{CH}=\begin{pmatrix}
    s_2 c_1 & s_2 s_1 & c_2 \\
    -c_1 c_2 s_\omega- s_1 c_\omega & c_1 c_\omega - s_1 c_2 s_\omega & s_2 s_\omega \\
    s_1 s_\omega - c_1 c_2 c_\omega & - c_1 s_\omega - s_1 c_2 c_\omega & s_2 c_\omega
\end{pmatrix},
\label{eq:3hdm_decoupling_matrix}
\end{equation}
with $s_1 \equiv \sin{\beta_1}$, $c_1 \equiv \cos{\beta_1}$,
$s_2 \equiv \sin{\beta_2}$, $c_2 \equiv \cos{\beta_2}$,
$s_\omega \equiv \sin{\omega}$, $c_\omega \equiv \cos{\omega}$.
As mentioned,
the angles $\beta_1$ and $\beta_2$ are determined solely by the vevs,
while the angle $\omega$ will depend on the quadratic and quartic
parameters in the S basis in some complicated fashion.

One final notational issue must be addressed.
If there are two doublets with the same group charge,
then any basis change among those two
doublets is allowed. For example,
the 3HDM with the $\mathbb{Z}_2$ symmetry
$S=\textrm{diag}(1,-1,-1)$ does not have a well defined
symmetry basis, since one can mix at will the last two scalars.
We will concentrate on models which do have a
\textit{well defined S basis},
such as the $\mathbb{Z}_2 \times \mathbb{Z}_2$ 3HDM 
generated by $S_1=\textrm{diag}(1,-1,1)$ and $S_2=\textrm{diag}(1,1,-1)$,
or the $\mathbb{Z}_3$ 3HDM generated by
$S=\textrm{diag}(1,\omega,\omega^2)$,
with $\omega^3 = 1$.

In summary, the S basis is useful to identify the set of independent parameters,
while the CH basis is useful to discuss decoupling.
Going from the former to the latter involves diagonalizing $N \times N$
(for the charged scalars) and $2N \times 2N$ matrices
(for the neutral scalars), and it is basically only manageable
for $N=2$ or other exceedingly simple cases.
We will now show that the converse procedure
of starting from quadratic parameters in the CH basis
and looking for generic properties of the quadratic parameters in
the S basis can lead to results valid for any $N$.

\section{\label{sec:decoupling}Decoupling or nondecoupling}


Consider a NHDM with some symmetry.
It may decouple to the SM in two ways:
\begin{itemize}
\item NHDM $\rightarrow$ SM, with one single scale
${\cal M}_{k \neq 1} = {\cal M} \gg v$;
\item NHDM $\rightarrow$ SM, with multiple scales
${\cal M}_k \gg  {\cal M}_j \gg v$.
\end{itemize}

\subsection{\label{subsec:singlescale}NHDM $\rightarrow$ SM: single scale}

Regarding the first possibility, we present our main

\textbf{Theorem 1:} Barring some zero vevs ($v_i=0$ for some $i$),
and/or very small mixing angles 
in the matrix $U^{CH}$,
then \textit{all $Y_{ij}^S$ must be present in order for a decoupling limit
NHDM $\rightarrow$ SM with one single scale to exist}.

In most cases, a symmetry $S$ forces some $Y_{ij}^S$ to vanish.
Our claim is that, under the conditions of the theorem,
such models will \textit{not} have a decoupling limit.
Conversely,
in order to keep the decoupling limit,
the way out is to break the symmetry softly by including
\textit{all} the quadratic terms.
This applies to any $N$ and any symmetry $S$.

The proof uses eq.~\eqref{eq:Nhdm_Sb_quadratic}.
Perturbativity implies that all $Z^{CH}_{ij,kl}$ must be smaller
than about $4 \pi$.
Thus, we conclude from
eq.~\eqref{eq:Nhdm_CHiggsb_stationary} that
each $\Omega^{CH}_{kl}$ must be smaller that $O(10) v^2$.
Taking all doublets to share a decoupling scale
\be
\mathcal{M}^2_{k \neq 1} = \mathcal{M}^2 \gg v^2,
\ee
eq.~\eqref{eq:Nhdm_Sb_quadratic} leads to
\begin{align}
\frac{Y^S_{ij}}{ \mathcal{M}^2} &\approx \sum^N_{k=2} U^{CH*}_{ki} U^{CH}_{kj}
+ O\left(\frac{v^2}{{\cal M}^2}\right)
\nonumber\\
& = \delta_{ij} - U^{CH*}_{1i} U^{CH}_{1j} + O\left(\frac{v^2}{{\cal M}^2}\right)
.
\label{main1}
\end{align}
Thus,
\begin{align}
\frac{Y^S_{ii}}{ \mathcal{M}^2} &\approx 1 - \frac{|v_i|^2}{v^2}
+ O\left(\frac{v^2}{{\cal M}^2}\right),
\label{main2}
\\
\frac{Y^S_{i\ j\neq i}}{ \mathcal{M}^2} &\approx - \frac{v_i v_j^*}{v^2}
+ O\left(\frac{v^2}{{\cal M}^2}\right).
\label{main3}
\end{align}
The right-hand side (RHS) of eq.~\eqref{main2}
must be of order one.
Similarly,
the RHS of eq.~\eqref{main3} must be of order one,
unless some vev is very small.

There are a few caveats.
First,
the theorem does not hold in directions with $v_i=0$;
ie, in inert models.
Second,
the theorem may also cease to hold if
there are vevs of order $v^2/{\cal M}$,
such that there is a cancellation between 
the two terms on the RHS  of eq.~\eqref{main3}.
In fact, in such circumstances,
it could even happen that some $Y_{ij}^S$ is exactly
zero.\footnote{See eq.~\eqref{ateda} below.}
Recall that,
according to eq.~\eqref{eq:Nhdm_CHiggsb_basis_change_2},
if $v_k \sim v^2/{\cal M}$ for some $k$,
then the corresponding angle in the first line entry
$U^{CH}_{1k}$ will be small,
of order $v/{\cal M}$.
Third, 
one could consider decoupling in multiple scales.

Notice that we followed an uncommon strategy.
Usually, one discusses the decoupling limit by 
starting from the restricted set of parameters in the S basis and then
finding how to rotate from this basis into a new basis.
For example,
Gunion and Haber \cite{Gunion:2002zf} start from the S basis of the
$\mathbb{Z}_2$ 2HDM,
construct the mass matrices in this basis, and then diagonalize them,
transforming from this basis into the mass eigenstates directly.
Later, the result was revisited by Bernon \textit{et. al.} \cite{Bernon:2015qea}
by starting again in the symmetry basis and going into the
Higgs basis.\footnote{There is no distinction between the
Higgs basis and the CH basis when $N=2$.
See ref.~\cite{Bento:2017eti} for details.}
Again, this requires that one minimizes the potential explicitly and
finds the matrix going from the S basis into the Higgs basis,
and then from this into the mass basis of the neutral scalars.
This is easy for $N=2$,
but unmanageable for larger $N$.
Here we follow the opposite strategy.
We write the S basis in term of the CH basis,
where decoupling is very easy to describe.

For our theorem, it was sufficient to use
the approximate results in eqs.~\eqref{main1}-\eqref{main3}.
To compare with exact results in the literature,
incorporating in the quartic couplings the
constraints from the symmetry $S$,
one can apply the following \textbf{Strategy:}
\begin{itemize}
\item find the constraints that the symmetry imposes on the quartic
couplings in the S basis, $Z^S_{mn,op}$;
\item write the quartic parameters in the CH basis,
$Z^{CH}_{ij,kl}$, using eq.~\eqref{eq:Nhdm_CHiggsb_quartic}.
This guarantees that the quartic parameters in the CH basis
already encode the constraints that the symmetry places
on the quartic couplings;
\item use eq.~\eqref{eq:Nhdm_Sb_quadratic}
to see how the quadratic parameters in the S basis
depend on the decoupling parameters $\mathcal{M}^2_{i}$.
\end{itemize}
Taking as an example the softly broken
$\mathcal{CP}$ conserving $\mathbb{Z}_2$ 2HDM,
\begin{align}
V_H &= Y_{11}^S |\Phi_1|^2 +  Y_{22}^S |\Phi_2|^2 
	+  Y_{12}^S \left[ \Phi_1^\dagger \Phi_2 +  \Phi_2^\dagger \Phi_1 \right]
\nonumber\\
&+ \lambda_1 |\Phi_1|^4 +  \lambda_2 |\Phi_2|^4 + \lambda_3  |\Phi_1|^2  |\Phi_2|^2
\nonumber\\
&+ \lambda_4  \left( \Phi_1^\dagger \Phi_2 \right) \left( \Phi_2^\dagger \Phi_1 \right)
+ \lambda_5 \left[ (\Phi_1^\dagger \Phi_2)^2 +  (\Phi_2^\dagger \Phi_1)^2 \right],
\end{align}
we find
\begin{equation}
Y^S_{12} =
- \frac{s_{2\beta}}{2}
\left\{
\mathcal{M}_2^2
+ \frac{v^2}{2}
	\left[
	s^2_{\beta}c^2_{\beta}\lambda_{12}
	+ \left( c^4_{\beta} + s^4_{\beta} \right)\lambda_{345}
	\right]
\right\},
\label{eq:2hdm_Higgsb_decoupling_quadratic}
\end{equation}
where $\tan{\beta}=v_2/v_1$,
$\lambda_{12} = \lambda_1+\lambda_2$,
and $\lambda_{345}=\lambda_3+\lambda_4+\lambda_5$.
One can explicitly see that with $\beta \neq 0,\, \pi/2$ it is not possible
to simultaneously have $\mathcal{M}_2 \gg v$ and $Y^S_{12}=0$.
We recover the result that the exact $\mathbb{Z}_2$ symmetric 2HDM does
not have a decoupling limit \cite{Gunion:2002zf,Bernon:2015qea}.
Moreover, if $\beta=0$, then we are in the case of the Inert Doublet Model,
which can indeed have a decoupling limit while $Y^S_{12}=0$.

\subsection{\label{subsec:multiplescales}NHDM $\rightarrow$ SM:
multiple scales}

We now consider the possibility that there are multiple
$Y_{kk}^{CH} \equiv {\cal M}_k^2 \gg v^2$ ($k \neq 1$) scales
in the decoupling NHDM $\rightarrow$ SM. For definiteness,
one could imagine that ${\cal M}_3 \gg {\cal M}_2 \gg v$.
We assume that the two scales are distinct, but also independent.
For example, we exclude the possibility that
${\cal M}_3 = 100\, {\cal M}_2$,
both growing towards very high values in proportion to each other.
We find the following

\textbf{Theorem 2:} Barring some zero vevs ($v_i=0$ for some $i$),
and/or very small mixing angles in the matrix $U^{CH}$,
in order for a multiple scale NHDM $\rightarrow$ SM decoupling to exist:
\begin{itemize}
\item[a)]
all $Y_{ii}^S$ must be present and large;
\item[b)] moreover, if
there are no
judicious cancellations,
then all $Y_{i\ j\neq i}^S$ must be present and with large magnitudes.
\end{itemize}

The proof of a) is very similar to that in the previous section.
Taking
\be
\mathcal{M}^2_{k \neq 1} \gg v^2,
\ee
eq.~\eqref{eq:Nhdm_Sb_quadratic} leads to
\begin{equation}
Y^S_{aa} \approx \sum^N_{k=2} |U^{CH}_{ka}|^2\  \mathcal{M}^2_k 
+ O(v^2).
\label{main21}
\end{equation}
For a fixed $a$,
$Y^S_{aa}$ can only be of order $v^2$ (or vanish)
if $U^{CH}_{ka}=0$ \textit{for all values of $k \neq 1$}.
But this is impossible.
Indeed, if it were true,
the unitarity condition
\be
1 = \sum_{k=1}^N |U^{CH}_{ka}|^2 = |U^{CH}_{1a}|^2 = \frac{|v_1|^2}{v^2}
\ee
would force $|v_1|=v$ and \textit{all} $v_{k \neq 1}=0$, contradicting our hypothesis.

To prove b) we start from
\begin{equation}
Y^S_{ab} \approx \sum^N_{k=2}  U^{CH*}_{ka} U^{CH}_{kb}\  \mathcal{M}^2_k
+ O(v^2).
\label{main22}
\end{equation}
We assume that there is no judicious cancellation among
large terms,
such as
\be
 U^{CH*}_{2a} U^{CH}_{2b}\  \mathcal{M}^2_2
+
 U^{CH*}_{3a} U^{CH}_{3b}\  \mathcal{M}^2_3
= 0.
\label{maybe_S}
\ee
Such a cancellation could occur because one is looking at a particularly
fine-tunned region of parameters.\footnote{For example,
in eq.~\eqref{eq:3hdm_SB_decoupling_12} below,
valid for the $\mathbb{Z}_2 \times \mathbb{Z}_2$ 3HDM,
one sees that $Y_{12}^S$ is proportional to
$\mathcal{M}^2_{3} - \mathcal{M}^2_{2}
+ \frac{v^2}{2} \left(\lambda _{13}-\lambda _{23} \right)$.
One could then choose the fine-tuned parameter region
$\mathcal{M}^2_{3} = \mathcal{M}^2_{2}
+ \frac{v^2}{2} \left(\lambda _{13}-\lambda _{23} \right)$,
thus allowing $Y_{12}^S=0$.
The same holds in eq.~\eqref{eq:3hdm_decoupling_second_12} below,
applicable to the $\mathbb{Z}_3$ 3HDM.
It is this type of situation, possible only when there are
two mass scales, that the ``no judicious cancellations''
requirement precludes.}
But it could also occur naturally due to the symmetry.
Recall that, given some non-Abelian symmetry,
one can diagonalize one generator, but not another.
The latter will necessarily impose relations between
parameters corresponding to different doublets and,
under those circumstances, an equation such as \eqref{maybe_S}
cannot be excluded a priori. The theorem b) applies when there
is no such cancellation.
In addition,
we are also excluding situations in which there are small entries
$U^{CH*}_{ka} U^{CH}_{kb}$,
such that the RHS of eq.~\eqref{main22} has some term in ${\cal M}_k^2$
exactly canceled by some term of order $v^2$.
Excluding such judicious cancellations,
for given $a$ and $b \neq a$,
$Y^S_{ab}$ in eq.~\eqref{main22} can only vanish if
all combinations $U^{CH*}_{ka} U^{CH}_{kb} = 0$
\textit{for all values of $k \neq 1$}.
But this is impossible.
Indeed, if it were true,
the unitarity condition
\be
0 = \sum_{k=1}^N U^{CH*}_{ka} U^{CH}_{kb}  = U^{CH*}_{1a} U^{CH}_{1b} 
= \frac{v_a^* v_b}{v^2}
\label{main23}
\ee
would force some $v_{i \neq 1}=0$, contradicting our hypothesis.


Notice that,
in both Theorem 1 and Theorem 2,
the result that $Y_{aa}^S$ must be present is not very useful in cases
in which all generators of the symmetry are simultaneously diagonalizable
and chosen to be represented as in eq.~\eqref{S-diagonal-1}.
Indeed, in such cases,
all $Y^S_{aa}$ are allowed by the symmetry from the start.
In contrast,
in cases where the generators cannot be simultaneously diagonalized,
such as the case mentioned in eq.~\eqref{ab},
then one or more $Y^S_{aa}$ might be precluded by the exact symmetry.
In that case, 
Theorem 1 and case a) of Theorem 2 assert that the
symmetry must be softly violated by $Y_{aa}^S \neq 0$.

\subsection{\label{subsec:onek}Decoupling of one doublet}

It is interesting that a strong result is possible,
even under the simple situation in which only one
doublet decouples.

\textbf{Theorem 3:} Barring some zero vevs ($v_i=0$ for some $i$),
for every doublet, $\Phi^{CH}_{k\neq 1}$, that decouples from the low energy
theory, and barring judicious cancellations,
at least three quadratic parameter in the Symmetry basis,
$Y^S_{ab \neq a}$ $Y^S_{aa}$ and $Y^S_{bb}$,
will depend on $\mathcal{M}^2_{k\neq 1}$.

The proof uses eq.~\eqref{eq:Nhdm_Sb_quadratic},
eq.~\eqref{eq:Nhdm_CHiggsb_basis_change_2},
and the fact that the matrix $U^{CH}$ is unitary.
The latter implies that the sum of the squares of the entries in
a given row (column) must be unity.
Since we are excluding judicious cancellations (that is,
proportionality among) $\mathcal{M}^2_{k\neq 1} $ for different $k$,
we can consider the impact of each one independently.
Imagine that for a given $k \neq 1$, $\mathcal{M}^2_{k\neq 1} \gg v^2$,
and consider the $k$-th row of $U^{CH}$. Since the sums of squares along
the row must add to one, there must be at least a column $a$ such that
$|U_{ka}| \neq 0$. However, we know that also $|U_{ka}| \neq 1$;
otherwise, considering now the sum of all squares in column $a$,
we would find $|U_{1a}|=0$. That,
according to eq.~\eqref{eq:Nhdm_CHiggsb_basis_change_2},
would imply $v_a=0$, violating our hypothesis.
So, there must be at least one other column $b$ such that
$|U_{kb}| \neq 0$.
But, looking back at eq.~\eqref{eq:Nhdm_Sb_quadratic} we see that,
as a result, $Y^S_{ab \neq a}$, $Y^S_{aa}$, and $Y^S_{bb}$ grow
with $\mathcal{M}^2_{k\neq 1}$.

It may help to visualize the argument made here to look back at
eq.~\eqref{eq:3hdm_decoupling_matrix}.
Imagine that we are taking $\mathcal{M}^2_2 \gg v^2$.
One can make $U^{CH}_{23} = 0$ by setting
$\omega = 0$; the matrix simplifies into
\begin{equation}
U^{CH}=\begin{pmatrix}
    s_2 c_1 & s_2 s_1 & c_2 \\
    - s_1  & c_1  & 0 \\
    - c_1 c_2 & \ \ - s_1 c_2  & \ \ s_2 
\end{pmatrix}.
\label{eq:3hdm_decoupling_matrix_omega_0}
\end{equation}
Then, according to eq.~\eqref{eq:Nhdm_Sb_quadratic},
none of $Y^S_{i3}$ and $Y^S_{3i}$ depend on
 $\mathcal{M}^2_2$.
Nevertheless,
both $U^{CH}_{21}$ and $U^{CH}_{22}$ must be nonvanishing,
or there would be one zero vev.
As a result,
even taking $\omega=0$,
$Y^S_{12}$, $Y^S_{11}$, and $Y^S_{22}$
will grow with $\mathcal{M}^2_2$.

\section{\label{sec:examples}Simple examples}

To illustrate both the application of our theorems (decoupling)
and some relevant violations of the assumptions (thus, nondecoupling
or decoupling using fine tuned regions with vevs of order $v^2/{\cal M}$),
we concentrate on 3HDM models with Abelian symmetries.
These have been classified in \cite{Ferreira:2008zy} and \cite{Ivanov:2011ae}.
The symmetries $\mathbb{U}(1) \times \mathbb{U}(1)$,
$\mathbb{U}(1) \times \mathbb{Z}_2$, $\mathbb{Z}_2 \times \mathbb{Z}_2$,
$\mathbb{Z}_3$, and $\mathbb{Z}_4$,
yield a well defined symmetry basis.
The corresponding potentials are shown in \ref{app:welldefined}.
All these symmetries allow for quadratic diagonal coefficients $Y^S_{ii}$
and preclude quadratic off-diagonal coefficients $Y^S_{ij \neq i}$.
When off-diagonal quadratic couplings $Y^S_{ij \neq i}$
are needed for decoupling, then one must either give up
decoupling or else break the symmetry softly.

\subsection{\label{subsec:Z2xZ2}The $\mathbb{Z}_2 \times \mathbb{Z}_2$ 3HDM}

This model was suggested by Weinberg \cite{Weinberg:1976hu} and explored
for spontaneous $\mathcal{CP}$ violation with three quark families
by Branco \cite{Branco:1979pv}.
The potential is written in eqs.~\eqref{eq:V0_potential_3hdm} and
\eqref{eq:exact_potential_z2xz2}.
We now follow the strategy outlined at the end of 
section~\ref{subsec:singlescale}.

To simplify the expressions, we take the vevs real and
use eq.~\eqref{eq:3hdm_decoupling_matrix} with
$\omega=0$,
$\cos{(\beta_2)} \approx 1-\beta^2_2/2$,
$\sin{(\beta_2)} \approx \beta_2$,
$\cos{(\beta_1)} \approx 1-\beta^2_1/2$, $\sin{(\beta_1)} \approx \beta_1$.
These conditions ensure that $v_3 \rightarrow v$ and that the
contributions for the $Y^S_{ij\neq i}$ due to
$\mathcal{M}^2_{k \neq 1} \gg v^2$ are suppressed.
Using eqs.~\eqref{eq:Nhdm_CHiggsb_quadratic},
\eqref{eq:Nhdm_CHiggsb_stationary},
and \eqref{eq:Nhdm_CHiggsb_quartic}, we find
\begin{align}
    Y^{CH}_{11}&=-v^2Z^{CH}_{11,11}
\nonumber\\
& \approx-v^2\left[\frac{\lambda_{33}}{2} + \beta _2^2\left(\lambda_{13}
+ \lambda^\prime_{13} + \lambda _{13,13} - \lambda_{33}\right)\right],\\
    Y^{CH}_{12}&=-v^2Z^{CH}_{12,11}
\nonumber\\
& \approx \frac{v^2}{2}\beta _1 \beta _2 \left(\lambda_{13}
+ \lambda^\prime_{13} + \lambda _{13,13}
\right.
\nonumber\\
& \hspace{11ex}\left.
- \lambda_{23} - \lambda^\prime_{23} - \lambda _{23,23} \right),\\
    Y^{CH}_{13}&=-v^2Z^{CH}_{13,11}
\nonumber\\
& \approx \frac{v^2}{2} \beta _2 \left(\lambda_{13} + \lambda^\prime_{13}
+ \lambda _{13,13} -\lambda_{33}\right),\\
Y^{CH}_{22}&={\cal M}_2^2, \\
    Y^{CH}_{23}&=-v^2Z^{CH}_{23,11} \approx \frac{v^2}{2} \beta _1 \left(\lambda_{23} - \lambda_{13}\right),\\
Y^{CH}_{33}&={\cal M}_3^2.
\end{align}
Substituting in eq.~\eqref{eq:Nhdm_CHiggsb_basis_change_quadratic},
we get the leading order terms
\begin{align}
Y^S_{11} & \approx \mathcal{M}^2_{3} + \beta _1^2 \left[ \mathcal{M}^2_{2} -\mathcal{M}^2_{3}
-v^2 \left(\lambda_{13}-\lambda _{23}\right)\right]
\nonumber\\
&
\hspace{2ex} -  \beta_2^2 \left[\mathcal{M}^2_{3} + v^2 \left(\lambda_{13} + \lambda^\prime_{13}
+ \lambda_{13,13} - \frac{\lambda_{33}}{2}\right) \right],
\label{eq:3hdm_SB_decoupling_11} \\
Y^S_{12} & \approx \beta _1\left[ \mathcal{M}^2_{3} - \mathcal{M}^2_{2}
+ \frac{v^2}{2} \left(\lambda _{13}-\lambda _{23} \right) \right],
\label{eq:3hdm_SB_decoupling_12} \\
Y^S_{13} & \approx -\beta _2 \left[ \mathcal{M}^2_{3} + \frac{v^2}{2}
\left(\lambda_{13}+\lambda^\prime_{13}+\lambda _{13,13}\right) \right],
\label{eq:3hdm_SB_decoupling_13} \\
Y^S_{22} & \approx \mathcal{M}^2_{2} + \beta^2 _1\left[ \mathcal{M}^2_{3} -\mathcal{M}^2_{2}
+ v^2 \left(\lambda _{13}-\lambda _{23}\right) \right],
\label{eq:3hdm_SB_decoupling_22} \\
Y^S_{23} & \approx -\beta _1 \beta _2  \left[ \mathcal{M}^2_{3} + 
\frac{v^2}{2} \left(\lambda_{13}+\lambda^\prime_{23}
+\lambda _{23,23}\right) \right],
\label{eq:3hdm_SB_decoupling_23} \\
Y^S_{33} & \approx \mathcal{M}^2_{3}\beta^2 _2 -\frac{v^2}{2} \lambda_{33} (1-\beta_2^2),
\label{eq:3hdm_SB_decoupling_33}
\end{align}

We start by noticing that,
as proved in section~\ref{subsec:onek},
$Y^S_{12}$, $Y^S_{11}$, and $Y^S_{22}$
grow with $\mathcal{M}^2_2$.

Let us now assume, as in section~\ref{subsec:multiplescales},
that there are two very different scales
${\cal M}_3 \gg {\cal M}_2 \gg v$ in the decoupling 3HDM $\rightarrow$ SM.
Then, 
all $Y^S_{ij}$ must be present.
However,
as we see in eq.~\eqref{eq:V0_potential_3hdm},
the $\mathbb{Z}_2 \times \mathbb{Z}_2$ symmetry precludes
the $Y^S_{ij \neq i}$ terms.
Thus,
in this region of the parameter space,
one either keeps the symmetry and loses the decoupling limit or, else,
one must break the symmetry softly with all possible terms.
It is true that one could (for example) make $Y^S_{13}$
in eq.~\eqref{eq:3hdm_SB_decoupling_13}  parametrically small
by choosing fine tuned regions with 
$\beta_2 \sim v^2/{\cal M}_3^2$.
Small but \textit{not zero}.
Again, 
in the particular region
$\omega = 0$, $\beta_1, \beta_2 \ll 1$,
for a 3HDM $\rightarrow$ SM with several scales
one must include \textit{all} $Y^S_{ij \neq i}$ terms to have a decoupling limit.
Notice that this holds despite the fact that in this particular
case we are even allowing small vevs.

Now, we notice that the leading term for 
$Y^S_{12}$ in eq.~\eqref{eq:3hdm_SB_decoupling_12} is proportional to
$\mathcal{M}^2_{3} - \mathcal{M}^2_{2}$.
So, we consider the situation where there is only one scale
$\mathcal{M}= \mathcal{M}^2_{3} = \mathcal{M}^2_{2} \gg v^2$
in the $\mathbb{Z}_2 \times \mathbb{Z}_2$ 3HDM $\rightarrow$ SM decoupling,
as discussed in section~\ref{subsec:singlescale}.
One finds
\begin{align}
Y^S_{11} & \approx \mathcal{M}^2
- \beta _1^2 v^2 \left(\lambda_{13}-\lambda _{23}\right)
\nonumber\\
&
\hspace{2ex} -  \beta_2^2 \left[\mathcal{M}^2
+ v^2 \left(\lambda_{13} + \lambda^\prime_{13}
+ \lambda_{13,13} - \frac{\lambda_{33}}{2}\right) \right],
\label{eq:3hdm_SB_decoupling_11_v2} \\
Y^S_{13} & \approx -\beta _2 \left[ \mathcal{M}^2 + \frac{v^2}{2}
\left(\lambda_{13}+\lambda^\prime_{13}+\lambda _{13,13}\right) \right],
\label{eq:3hdm_SB_decoupling_13_v2} \\
Y^S_{22} & \approx \mathcal{M}^2
+ \beta^2 _1 v^2 \left(\lambda _{13}-\lambda _{23}\right),
\label{eq:3hdm_SB_decoupling_22_v2} \\
Y^S_{23} & \approx -\beta _1 \beta _2  \left[ \mathcal{M}^2 + 
\frac{v^2}{2} \left(\lambda_{13}+\lambda^\prime_{23}
+\lambda _{23,23}\right) \right],
\label{eq:3hdm_SB_decoupling_23_v2} \\
Y^S_{33} & \approx \mathcal{M}^2 \beta^2 _2 -\frac{v^2}{2} \lambda_{33} (1-\beta_2^2),
\label{eq:3hdm_SB_decoupling_33_v2}
\end{align}
and, taking $Y^S_{12}$ to higher order in $\beta$'s,
\be
Y^S_{12}  \approx -\beta _1  \left[ \mathcal{M}^2 \beta^2 _2 - \frac{v^2}{2} \left(\lambda_{13}-\lambda_{23}\right)\right].
\label{ateda}
\ee
Now it is possible to make $Y^S_{12}$
\textit{exactly zero},
by choosing fine tuned regions of
$\lambda_{13}-\lambda_{23}$, ${\cal M}$ (very large) and
$\beta_2$ (correspondingly very small),
such that the RHS of eq.~\eqref{ateda} vanishes.
Notice that,
according to eq.~\eqref{vevs_123},
$\beta_2 \rightarrow 0$ implies $v_1 \rightarrow 0$ and
$v_2 \rightarrow 0$,
in contradiction with the conditions of Theorem 1.
For the general NHDM, we do not know how many $Y^S_{ij\neq i}$ need to be
included in order to attain this parametric decoupling with
$\mathcal{M}_{k \neq 1}=\mathcal{M} \gg v$.
Nevertheless, we suspect that it may be possible to exclude
$Y^S_{ab\neq a}$ when both $v_a \rightarrow 0$ and $v_b \rightarrow 0$;
granted, using a very fined tuned choice of parameters.

Although simplified by taking $\omega=0$ and small $\beta_1$ and
$\beta_2$,
eqs.~\eqref{eq:3hdm_SB_decoupling_11}-\eqref{eq:3hdm_SB_decoupling_33}
and their special
case \eqref{eq:3hdm_SB_decoupling_11_v2}-\eqref{ateda}
illustrate both the theorems and their caveats.
Eqs. \eqref{eq:3hdm_SB_decoupling_11_v2}-\eqref{ateda} show that,
barring $\beta$'s decreasing as some power of $v/{\cal M}$,
decoupling implies that all quadratic couplings must be present,
in accordance with Theorem 1.
Similarly,
eqs.~\eqref{eq:3hdm_SB_decoupling_11}-\eqref{eq:3hdm_SB_decoupling_33}
illustrate Theorem 2.
Finally,
looking at 
eqs.~\eqref{eq:3hdm_SB_decoupling_11}-\eqref{eq:3hdm_SB_decoupling_33}
by taking only ${\cal M}_3^2 \gg v^2$ we see that all quadratic
couplings must be present in the S basis.
Moreover, 
taking only ${\cal M}_2^2 \gg v^2$ we see that some quadratic
couplings may be absent, but that there are at least three which must be
present.\footnote{It was to illustrate the fact that we could allow some
(but not all) quadratic parameters to be absent in some regions
of parameter space, that we have taken $\omega=0$.}
This illustrates Theorem 3.

This section also illustrates the caveats in the theorems.
For example,
by taking small angles $\omega=0$ and very fine tuned
regions of parameter space where the vevs decrease with
the decoupling scale (eg, $\beta_2 \sim v/{\cal M}$),
eq.~\eqref{ateda} shows that we can set $Y^S_{12}=0$
in the 3HDM $\rightarrow$ SM decoupling with one single scale.
Eqs.~\eqref{eq:3hdm_SB_decoupling_11}-\eqref{eq:3hdm_SB_decoupling_33}
show that that is not possible with multiple independent scales
${\cal M}_3^2 \gg {\cal M}_2^2\gg v^2$ for this fine tuned region of parameter
space where $\omega=0$ while $\beta_1$ and $\beta_2$ are small.

However,
in other fine tuned regions of the $\mathbb{Z}_2 \times \mathbb{Z}_2$ 3HDM
parameter space, it is indeed
possible to remove some $Y^S_{ab}=0$ for $a\neq b$,
even with multiple energy scales.
Consider the situation with
$\omega=0$, $\beta_2=\pi/2-\epsilon$,
$\cos{(\beta_2)} \approx \epsilon$,
$\sin{(\beta_2)} \approx 1-\epsilon^2/2$,
$\cos{(\beta_1)} \approx 1-\beta^2_1/2$,
and $\sin{(\beta_1)} \approx \beta_1$.
Using eqs.~\eqref{eq:Nhdm_CHiggsb_quadratic},
\eqref{eq:Nhdm_CHiggsb_stationary},
and \eqref{eq:Nhdm_CHiggsb_quartic}, we find
$Y^{CH}_{ij}$,
%
which we then substitute in
eq.~\eqref{eq:Nhdm_CHiggsb_basis_change_quadratic}.
The leading order terms are
\begin{align}
Y^S_{11} &\approx -\frac{1}{2} v^2 \lambda _{11}(1-\epsilon^2-\beta^2_1)
+\epsilon ^2 \mathcal{M}^2_3 +\beta _1^2\mathcal{M}^2_2, \\
Y^S_{12}&\approx \beta _1 \left[\mathcal{M}^2_3\epsilon ^2 -\mathcal{M}^2_2
-\frac{v^2}{2} \left(\lambda _{12}+\lambda^\prime _{12}+\lambda _{12,12}\right)\right], \\
Y^S_{13}&\approx -\epsilon  \left[\mathcal{M}^2_3 + \frac{v^2}{2}
\left(\lambda _{13}+\lambda^\prime _{13}+\lambda_{13,13}\right)\right], \\
Y^S_{22} & \approx \mathcal{M}^2_2(1-\beta_1^2) + \beta _1^2 v^2
\left(\frac{\lambda _{11}}{2}-\lambda _{12}-\lambda^\prime_{12}
-\lambda_{12,12}\right),\\
Y^S_{23}& \approx  - \epsilon\beta _1\left[ \mathcal{M}^2_3 +
\frac{v^2}{2} \left(\lambda _{13}+\lambda^\prime_{23}+\lambda _{23,23}\right) \right], \\
Y^S_{33} & \approx \mathcal{M}^2_3 (1 - \epsilon^2) +
\epsilon ^2 v^2 \left(\frac{\lambda _{11}}{2}
- \lambda _{13} - \lambda^\prime _{13} - \lambda _{13,13}\right).
\end{align}
By choosing a fined tuned region of the parameter space
where
\begin{equation}
    \mathcal{M}^2_3 = \frac{1}{\epsilon ^2}\left[\mathcal{M}^2_2
    +\frac{v^2}{2} \left(\lambda _{12}+\lambda^\prime _{12}
    +\lambda _{12,12}\right)\right],
\label{fine_tuned_Z2Z2_2}
\end{equation}
one can now set $Y^S_{12}=0$.
Notice that in this situation it is possible to set $Y^S_{12}=0$
as $\mathcal{M}^2_3$ takes increasingly larger values,
regardless of $\mathcal{M}^2_2$,
as long as one remains in the fine tuned region of
eq.~\eqref{fine_tuned_Z2Z2_2}.
However, one can never set $Y^S_{13}=0$ for this region of parameter space.

\subsection{The $\mathbb{Z}_3$ 3HDM}

To study the decoupling properties of the $\mathbb{Z}_3$ 3HDM,
we follow the strategy outlined at the end of 
section~\ref{subsec:singlescale}.
To simplify the expressions, we take again the vevs real,
use eq.~\eqref{eq:3hdm_decoupling_matrix} with
$\omega=0$,
and expand in $\beta_1$ and $\beta_2$.
These conditions ensure that $v_3 \rightarrow v$ and that the
contributions for the $Y^S_{ij\neq i}$ due to
$\mathcal{M}^2_{k \neq 1} \gg v^2$ are suppressed.
Using eqs.~\eqref{eq:Nhdm_CHiggsb_quadratic},
\eqref{eq:Nhdm_CHiggsb_stationary},
and \eqref{eq:Nhdm_CHiggsb_quartic}, we find
$Y^{CH}_{ij}$,
which we then substitute in
eq.~\eqref{eq:Nhdm_CHiggsb_basis_change_quadratic}.
The leading order terms are
\begin{align}
Y^S_{11} &
\approx \mathcal{M}^2_3 + \beta^2_1\left[ \mathcal{M}^2_2
- v^2\left(\lambda_{13}-\lambda_{23}\right)\right]
\nonumber\\
&\hspace{4mm}- v^2 \beta_2\left[\beta _2 \left( \lambda_{13}+ \lambda^\prime_{13}-\frac{\lambda_{33}}{2}\right) - \beta _1 \lambda _{13,12}\right],
\\
Y^S_{12} &
\approx \beta _1 \left[\mathcal{M}^2_{3} -\mathcal{M}^2_{2}
+ \frac{v^2}{2}\left( \lambda_{13} - \lambda_{23} -\beta _2 \lambda _{12,32}
\right)\right]
\nonumber\\
&\hspace{4mm}-\frac{v^2}{2} \beta _2 \lambda _{13,12},
\\
Y^S_{13} &
\approx -\beta _2 \left[ \mathcal{M}^2_{3}
+ \frac{v^2}{2}\left( \lambda_{13} + \lambda^\prime_{13}
+ \beta _1 \lambda _{13,23} \right)\right],
\\
Y^S_{22} &
\approx \mathcal{M}^2_{2} + \beta _1 \left[
\beta_1 \mathcal{M}^2_{3} + v^2 \beta_1\left(\lambda_{13}-\lambda_{23}\right)
\phantom{\frac{v^2}{2}}
\right.
\nonumber\\
&\hspace{1mm} \left.
\phantom{\frac{v^2}{2}} - v^2 \beta _2 \lambda _{13,12}\right], 
\\
Y^S_{23} & \approx -\beta _2 \left[\beta _1 \mathcal{M}^2_{3}
+  \frac{v^2}{2}\beta _1 \left(\lambda_{13}+\lambda^\prime_{23}\right)
+\frac{v^2}{2}\lambda _{13,23}\right],
\\
Y^S_{33} & \approx \beta _2^2 \mathcal{M}^2_{3} - \frac{1}{2} v^2 \lambda_{33}.
\end{align}
Here one can see that one can set $Y^S_{12}=Y^S_{23}=0$,
by choosing a fine tuned region of parameter space with decoupling energy
scales given to order $\beta_i v^2$ by
\begin{align}
   \mathcal{M}^2_{2} -  \mathcal{M}^2_{3}
   & \approx -\frac{v^2}{2} \frac{\beta_2}{\beta_1}\lambda _{13,12}
   \nonumber\\
	&\hspace{4mm} + \frac{v^2}{2}\left(\lambda_{13} - \lambda_{23} -\beta_2
   \lambda _{12,32} \right),
\label{eq:3hdm_decoupling_second_12}\\
    \mathcal{M}^2_{3}
    & \approx - \frac{v^2}{2}\frac{1}{\beta_1}\lambda _{13,23}
    - \frac{v^2}{2}\left(\lambda_{13}+\lambda^\prime_{23}\right).
\label{eq:3hdm_decoupling_second_23}
\end{align}
One concludes that when $\beta_1 \rightarrow 0 $ with
$\beta_2$ fixed, the decoupling energy scales can
still be larger than the electroweak scale without including
all quadratic parameters.
Through this procedure, it is possible to decouple a
3HDM $\rightarrow$ SM with two scales which
become larger and farther apart as $\beta_1 \rightarrow 0$.

\subsection{Lessons from the stationarity equations}

In all cases illustrated in section~\ref{subsec:Z2xZ2},
the $\mathbb{Z}_2 \times \mathbb{Z}_2$ 3HDM can have decoupling
if and only if there is at least some off-diagonal term
breaking the symmetry softly.
It is easy to see that, indeed,
this is a general feature of the $\mathbb{Z}_2 \times \mathbb{Z}_2$ 3HDM 
by writing the stationarity equations in the symmetry basis:
\begin{align}
    m^2_{11}&=-\frac{1}{2}\left[v_1^2 \lambda _{11}
    + v_2^2 \left(\lambda_{12}+\lambda^\prime_{12}+\lambda _{12,12}\right)
    \right.
    \nonumber\\
    & \hspace{4mm} \left.
    +v_3^2 \left(\lambda_{13}+\lambda^\prime_{13}+\lambda _{13,13}\right)\right],
    \nonumber\\
    m^2_{22}&=-\frac{1}{2}\left[v_1^2 \left(\lambda_{12}
    +\lambda^\prime_{12}+\lambda _{12,12}\right) + v_2^2 \lambda _{22}
    \right.
    \nonumber\\
    & \hspace{4mm} \left.
    + v_3^2 \left(\lambda_{23}+\lambda^\prime_{23}+\lambda _{23,23}\right)\right],
    \nonumber\\
    m^2_{33}&=-\frac{1}{2}\left[v_1^2 \left(\lambda_{13}
    +\lambda^\prime_{13}+\lambda _{13,13}\right)
    \right.
    \nonumber\\
    & \hspace{4mm} \left.
    +v_2^2 \left(\lambda_{23}+\lambda^\prime_{23}+\lambda _{23,23}\right)
    +v_3^2 \lambda _{33}\right].
    \label{SC:Z2xZ2}
\end{align}
We see that,
in the exact $\mathbb{Z}_2 \times \mathbb{Z}_2$ 3HDM 
(that is, with no $Y^S_{ij\neq i}$ terms),
all diagonal quadratic couplings $Y^S_{ii}=m_{ii}^2$ are or
order $v^2$, and thus there is no decoupling limit.
Hence, the stationarity conditions of the exact
$\mathbb{Z}_2 \times \mathbb{Z}_2$ 3HDM 
in the symmetry basis are enough to see that 
one can only make the diagonal $Y^S_{ii}=m_{ii}^2$ large
by including some off diagonal $Y^S_{ij\neq i}$ term.

This is no longer the case for the
$\mathbb{Z}_3$ 3HDM, whose potential 
is written in eqs.~\eqref{eq:V0_potential_3hdm} and
\eqref{eq:exact_potential_z3}.
Indeed,
the stationarity conditions for the
real $\mathbb{Z}_3$ 3HDM,
\begin{align}
    m^2_{11}&= -\frac{1}{2}\left[v_1^2 \lambda _{11}
    + v_2^2 \left(\lambda_{12} + \lambda^\prime_{12}
    + \frac{v_3}{v_1}\lambda _{12,32} \right)
    \right.
    \nonumber\\
    & \hspace{4mm}  \left.
    +\, v_3^2 \left(\lambda_{13} + \lambda^\prime_{13}
    + \frac{v_2}{v_1}\lambda _{13,23}\right)
    + 2 v_2 v_3\lambda _{13,12}\right],
    \nonumber\\
    m^2_{22}&= -\frac{1}{2}\left[ v^2_1\left( \lambda_{12}
    + \lambda^\prime_{12} +\frac{v_3}{v_2}\lambda _{13,12} \right)
    + v^2_2\lambda _{22}
    \right.
    \nonumber\\
    & \hspace{4mm} \left.
    +\, v^2_3\left( \lambda_{23} + \lambda^\prime_{23}
    + \frac{v_1}{v_2}\lambda_{13,23} \right) + 2v_1 v_3\lambda _{12,32} \right],
    \nonumber\\
    m^2_{33}&= -\frac{1}{2}\left[v_1^2 \left( \lambda_{13} + \lambda^\prime_{13}
    + \frac{v_2}{v_3}\lambda _{13,12}\right)
    \right.
    \nonumber\\
    & \hspace{4mm} 
    +\, v_2^2\left( \lambda_{23}+\lambda^\prime_{23} + \frac{v_1}{v_3}
    \lambda _{12,32}\right)
    \nonumber\\
    & \hspace{4mm} \left.
    +\, v_3^2 \lambda _{33} + 2 v_1 v_2\lambda _{13,23}\right],
    \label{SC:Z3}
\end{align}
contain ratios of vevs.
To be specific, take the expression for $m_{22}^2$.
Now one can have $m^2_{22} \gg v^2$ by taking $v_2 \rightarrow 0$ and
$v_1,\, v_3 \not\rightarrow 0 $.

It is interesting to interpret the 
difference between the stationarity conditions in this case,
eqs.~\eqref{SC:Z3}, and in the previous case,
eqs.~\eqref{SC:Z2xZ2},
in the following way.
The decoupling limit is easy to interpret in the CH basis.
It says that two parameters must become very large.
And, using eq.~\eqref{eq:Nhdm_CHiggsb_basis_change_quadratic}
and the fact that all $|U^{CH}_{ij}|<1$,
we conclude that some entries of $Y^S$ must be large.
Now,
the stationarity conditions of the general NHDM in the S basis
are
\begin{equation}
Y^S_{ii}=-\sum^N_{j=1, j\neq i}\left( Y^S_{ij}\frac{v_j}{v_i}
+ Z^S_{ij,kl}\frac{v^*_k v_l v_j}{v_i}\right),
\label{eq:Nhdm_Sb_stationary}
\end{equation}
Eq.~\eqref{eq:Nhdm_Sb_stationary} shows that there are two ways
to make the diagonal elements $Y^S_{ii}$ large.
First,
one can make the off-diagonal elements in the first term on the
RHS of eq.~\eqref{eq:Nhdm_Sb_stationary} large
\begin{equation}
Y^S_{ii}+\sum^N_{j=1, j\neq i}\left( Y^S_{ij}\frac{v_j}{v_i}\right)
= O(v^2).
\end{equation}
Or, one can use the ratios of vevs in the second term on the
RHS of eq.~\eqref{eq:Nhdm_Sb_stationary} and make those large
\begin{equation}
\left|Z^S_{ij,kl}\frac{v^*_k v_l v_j}{v_i}\right| \gg v^2 \quad \text{(no sum)}.
\end{equation}

We conjecture that this difference might have a physical impact.
We have found numerical regions of parameter space where an exact
$\mathbb{Z}_3$ 3HDM (with $Y_{ab}^S=0$ for \textit{all} $a\neq b$)
could decouple one scalar, leaving a 2HDM.
We have found no such situation for an exact 
$\mathbb{Z}_2 \times \mathbb{Z}_2$ 3HDM.

\section{\label{sec:cpv}A note on  $\mathcal{CP}$ violation}

The examples above focused on models with real vevs in a
real symmetry basis.
Then $\mathcal{CP}$ is conserved, and only the magnitudes of
the quadratic parameters are relevant for decoupling.
The same type of arguments can be used to study cases
where there is $\mathcal{CP}$ violation.
This is beyond the scope of this article, but we make
a few simple comments here.

Let us consider cases in which the potential is invariant under
the canonical definition of  $\mathcal{CP}$:
\be
\Phi_k \rightarrow \Phi_k^\ast.
\label{canonical_CP}
\ee
Invariance of the potential \eqref{eq:Nhdm_potential} under this
definition of $\mathcal{CP}$ implies that
\be
Y_{ij} = Y_{ij}^\ast,\ \ 
Z_{ij,kl} = Z_{ij,kl}^\ast,
\ee
and all coefficients are real.
Then one may have spontaneous $\mathcal{CP}$ violation if some
vevs have a relative phase.
Under this definition of $\mathcal{CP}$,
the fields $\rho_k$ and $\chi_k$ in
eq.~\eqref{eq:expansion_fields} are $\mathcal{CP}$-even
and $\mathcal{CP}$-odd,
respectively.
In that case,  $\mathcal{CP}$ violation in scalar-pseudoscalar mixing
is described by 
$\left(M^2_{\rho\chi}\right)_{ij}$ in eq.~\eqref{eq:Nhdm_mass_matrix_RI_v2},
which, since it is at most of order $v^2$,
becomes irrelevant as all charged Higgs become very massive.

There are several issues that complicate a general analysis.
First,
the definition of $\mathcal{CP}$ changes with basis
transformations.
For example, it is true that
under the $\mathcal{CP}$ transformation \eqref{canonical_CP},
$\rho_k$ and $\chi_k$ in
eq.~\eqref{eq:expansion_fields} are $\mathcal{CP}$-even
and $\mathcal{CP}$-odd.
However,
the simple rephasing
$\Phi_k \rightarrow \Phi_k^{\prime}= i \Phi_k$
means that, under the same definition of 
$\mathcal{CP}$,
it is now $\chi_k^\prime$ which is $\mathcal{CP}$-even
and $\rho_k^\prime$ which is $\mathcal{CP}$-odd.\footnote{In fact,
it was this precise problem which prompted the study of
basis invariants quantities in the scalar sector in
\cite{Lavoura:1994fv} and \cite{Botella:1994cs},
later explored extensively in
\cite{Branco:1999fs,Davidson:2005cw,Gunion:2005ja,Haber:2006ue,Haber:2010bw}.}
Second,
when searching for $\mathcal{CP}$ violation
one must study all possible definitions of
$\mathcal{CP}$ -- for an introduction to these problems,
see for example ref.~\cite{Branco:1999fs}.
Finally,
one may even have the odd situation that there is 
$\mathcal{CP}$ conservation even though there are irremovable
complex couplings in the scalar potential.
The first and simplest example is the so-called
CP4 3HDM \cite{Ivanov:2015mwl}.
For these reasons,
a complete analysis of the relation between
decoupling and $\mathcal{CP}$ violation
(explicit or spontaneous) lies beyond the scope of this article.

Still, the result discussed below eq.~\eqref{eq:calMi}
is completely general.
Indeed,
if all charged Higgs become very massive,
then eqs.~\eqref{MR_CH}-\eqref{MRI_CH}
imply that
$M_{\rho\rho}^2 $ and $M_{\chi\chi}^2 $ become very large and almost diagonal,
while $M_{\rho\chi}^2 $ remains of (small) order $v^2$.
Thus, all $\mathcal{CP}$ violation in scalar-pseudoscalar mixing vanishes,
regardless of the precise details of $\mathcal{CP}$.

We can also recover a recent result on the 2HDM \cite{Nebot:2019lzf},
because the model is very simple.
Consider a 2HDM with a softly broken $\mathbb{Z}_2$ symmetry and complex vevs.
As noted, the symmetry basis is defined up to rephasing of
its doublets.
Then one can without loss of generality choose $\lambda_5$ to be real.
In this basis $m^2_{12}$ can either be real or complex.
If it is real, there can be \cite{Branco:1985aq} (or not) spontaneous
$\mathcal{CP}$ violation,
and if it is complex, there is explicit $\mathcal{CP}$ violation.\footnote{This holds,
except in an exceptional region of parameter space \cite{2appear}.}
By parameterising the vevs as $v_1=v c_\beta$ and $v_2= vs_\beta e^{i\theta}$,
the transformation matrix from the S basis
with real $\lambda_5$, to the Higgs basis is given by,
\begin{equation}
    \begin{pmatrix}
    \Phi^{H}_1 \\ \Phi^{H}_2
    \end{pmatrix} =\frac{1}{v}\begin{pmatrix}
    v_1 & v^*_2 \\ -v^*_2 & v_1
    \end{pmatrix}
    \begin{pmatrix}
    \Phi_1 \\ \Phi_2
    \end{pmatrix}.
\end{equation}

To obtain the decoupling limit conditions we repeat our procedure
of setting $Y^{CH} \equiv \mathcal{M}^2_2$
and writing $Y^{CH}_{11}$, $Y^{CH}_{12}$ as a function
of the S basis quartic parameters.
Then we write the S basis quadratic parameters as
a function of the CH basis quadratic parameters.
Note that both $Y^{CH}_{12}$ and $U^{CH}$ are now complex.
Then,
\begin{equation}
Y^S=\mathcal{M}_{2}^2\begin{pmatrix}
 s^2_\beta        & -c_\beta s_\beta e^{-i\theta} \\
 -c_\beta s_\beta e^{i\theta} & c^2_\beta \end{pmatrix}
 + O\left(v^2\right),
\end{equation}
and the soft breaking term will be given by
\begin{equation}
    Y^S_{12}=-\frac{s_{2\beta}}{2}\left[
    \mathcal{M}^2_2 + O(v^2) \right]e^{-i\theta}.
\end{equation}
Here, one can see that it is not possible to have a decoupling limit
and spontaneous $\mathcal{CP}$ violation in a 2HDM with a
$\mathbb{Z}_2$ symmetry softly broken (by a real parameter).
$Y^S_{12}$ must be complex for a decoupling to exist, which in
turn explicitly breaks the $\mathcal{CP}$ symmetry.

\section{\label{sec:hypothesis}On the hypothesis of the theorems}

The previous theorems in this article assumed hypothesis ensuring that
the vevs are neither zero nor very small.
Without this, the various scenarios must be analyzed on a case-by-case basis.
But there is one further situation in which one can make a
statement valid for any NHDM.

\textbf{Theorem 4:} If the symmetry imposes $Y_{ab}^S=0$ for all $a \neq b$,
and all $Y_{aa}^S$ ($a \neq 1$) are independent of $Y_{11}^S$,
then the inert vacuum $v_1=v$, $v_{k\neq 1}=0$ allows for a decoupling limit.

Indeed, when $v_1=v$ and $v_{k\neq 1}=0$, the stationary conditions do not impose any restriction on the values of $Y^S_{kk}$ for all $k\neq 1$.
Moreover,
we are assuming that $Y_{kk}^S$ ($k \neq 1$) do not get
restricted by $Y_{11}^S$.\footnote{For example,
in the CP2, CP3 and $U(2)$ 2HDM cases discussed below,
$m_{22}^2=m_{11}^2$.
Combined with the fact that the stationarity equation
for the inert vacuum,
$2 m_{11}^2 = - \lambda_1 v^2$,
both are forced to be of order $v^2$ and there is no
parameter to drive decoupling.
This explains the ``$Y_{aa}^S$ ($a \neq 1$) are independent of
$Y_{11}^S$'' caveat of the theorem.}
Finally,
if $Y_{ab}^S=0$ for all $a \neq b$,
then  $Y^S_{kk} \approx Y^{CH}_{kk}= \mathcal{M}^2_k$
as $Y^S_{kk}\rightarrow \infty$,
and there is a valid decoupling limit.

This illustrates the difficulty one has with vanishing vevs; one must
then look in detail at the stationarity conditions to see which
vacuum solutions are valid and then find the various physical masses
for that specific vacuum.
In addition, some vacua may be solutions of the stationarity
equations but never constitute a valid global minimum for the
potential.
There is no known solution for these problems in a general NHDM.

Indeed, even simple questions in 3HDM have no known answer.
For example,
the full list of discrete symmetry
groups allowed in the 3HDM scalar sector was presented in
\cite{Ivanov:2012ry,Ivanov:2012fp}.
But this has never been attempted for the 4HDM and,
even for the 3HDM, there is still no complete description
of global minima or bounded from below conditions.
For example,
the necessary and sufficient bounded from below conditions
for the famous $Z_2 \times Z_2$ 3HDM
Weinberg model of 1976 \cite{Weinberg:1976hu}
are not yet known fully,
as explained recently in 
\cite{Faro:2019vcd}.

These difficulties highlight the power of the results presented here;
under the assumed hypothesis, they are valid in complete generality
for any NHDM. The difficulties also explain why we cannot give a
full answer to the issue we turn to next.

As noted at the end of section~\ref{subsec:singlescale},
the exact $Z_2$ 2HDM with the $Z_2$ symmetry breaking
vev $v_1 \neq 0$ and $v_2 \neq 0$ does
not have a decoupling limit.
In contrast,
the exact $Z_2$ 2HDM with the $Z_2$ symmetry preserving
vev $(v_1, v_2)=(v,0)$ -- the so-called Inert Doublet Model --
does have a decoupling limit.
Similarly,
at the end of the previous section~\ref{sec:cpv},
we recovered the result of \cite{Nebot:2019lzf}:
the exact CP conserving 2HDM with the CP violating vev
does not have a decoupling limit;
the exact CP conserving 2HDM with the CP conserving vev
does have a decoupling limit.
One may wonder whether this is a remark of
general validity.\footnote{We are very grateful to the anonymous referee
who asked the question which led to this section.}
That is, whether a symmetric NHDM will have a decoupling limit
if and only if the vacuum does not violate the symmetry.
Or, maybe, in which set of cases could such a statement hold.
The difficulties mentioned above explain why it is not easy to
find the answer to this question in the most general NHDM.

But one can say something within the 2HDM,
since all symmetries are known
and there are relatively few parameters.
The set of all possible symmetries which can be implemented in
the 2HDM was determined in \cite{Ivanov:2007de},
and further explained in \cite{Ferreira:2009wh}.
They were dubbed in \cite{Ferreira:2009wh} as $Z_2$, $U(1)$, $U(2)$,
CP1, CP2, and CP3.\footnote{The ``CP conserving 2HDM''
mentioned in the previous paragraph,
really means the most general CP conserving 2HDM, dubbed CP1
in \cite{Ferreira:2009wh}.}

The $U(1)$ 2HDM can be found from the $Z_2$ 2HDM by setting $\lambda_5=0$.
Here, the vev $v_1 \neq 0$ and $v_2 \neq 0$ breaks the symmetry.
And, since it is a continuous symmetry,
it implies that there is a massless Goldstone boson;
in the usual notation $m_A=0$.
This is as serious a nondecoupling limit as one could have.
In contrast,
the vev $(v_1, v_2)=(v,0)$ does not break $U(1)$ and
does have a decoupling limit,
with the masses of the new scalar $m_H$,
the new pseudoscalar $m_A$,
and the new charged scalar $m_{H^\pm}$ driven by $m_{22}^2$
(which is otherwise unconstrained).
For such an inert vacuum,
the only consequence of the increased $U(1)$ symmetry with respect to
the inert $Z_2$ model is that the neutral pseudoscalar 
becomes degenerate with one of the neutral scalars: $m_A=m_H$.

All nonzero vacua violate CP2, CP3 and $U(2)$.
Thus,
Theorem 1 states that there will be nondecoupling if the vacuum
has $v_1 \neq 0$ and $v_2 \neq 0$.
What happens in the inert minima $(v_1,v_2)=(v,0)$?
we start by noting that,
in the CP2, CP3 and $U(2)$ 2HDM, a basis can be chosen
where the parameters obey
$m_{22}^2 = m_{11}^2$, $m_{12}^2=0$, $\lambda_6=\lambda_7=0$.\footnote{This
is more difficult to see in the CP2 2HDM, but it is shown
in \cite{Davidson:2005cw,Maniatis:2006fs,Ivanov:2007de}.}
Moreover, for these parameters, the stationarity condition
forces $2 m_{11}^2 = - \lambda_1 v^2$.
Since $m_{22}^2=m_{11}^2$ and $m_{12}^2=0$,
there is no quadratic term to drive the decoupling.
Therefore, the exact CP2, CP3 and $U(2)$ 2HDM
will always have nondecoupling.

We have performed a complete analysis of all
symmetric 2HDM, where indeed
a symmetric 2HDM will have a decoupling limit
if and only if the vacuum does not violate the symmetry.
One might conjecture whether this is valid for all symmetric NHDM.
Lacking a classification of all symmetries in NHDM,
we have found no way to prove or disprove this conjecture.

\section{\label{sec:conclusions}Conclusions}

We have investigated the situations under which a generic NHDM might have
a decoupling limit.
This is important for model building, since one might wish to
gain intuition on a complicated model by studying it analytically
when it decouples into a simpler theory.
It is also very convenient when simulating numerically
a complicated model, by debugging with simulations of numerical limits
when it reduces effectively to a simpler model.

We have found that,
under the assumptions that the vevs and the mixing angles are not
parametrically small, \textit{all} quadratic couplings must be included
in order for decoupling to occur.
For the most part,
this means that either one has nondecoupling or, else,
one must break the symmetry softly.
In addition, we showed that there is no
$\mathcal{CP}$ violation in scalar-pseudoscalar mixing as the charged Higgs
masses approach decoupling.

Our theorems were illustrated with a number of special examples.
These were also used to explore violations of the assumptions
and to discuss what form of decoupling can occur in the
absence of some $Y^S_{i j \neq i}$, as long as one goes
to a very fine tuned region of parameter space,
where some vevs decrease as $v^2/{\cal M}$.


\begin{acknowledgements}
We are grateful to Miguel Bento, Igor Ivanov, and Patricia Conde Mu\'{\i}\~{n}o
for discussions,
and to Miguel Nebot for this and for commenting on the manuscript.
This work is supported in part
by the Portuguese \textit{Funda\c{c}\~{a}o para a Ci\^{e}ncia e Tecnologia}
(FCT) under contracts UIDB/00777/2020, UIDP/00777/2020,
CERN/FIS-PAR/0004/2017, and PTDC/FIS-PAR/29436/2017.
\end{acknowledgements}

\appendix

\section{Projective special unitary group}
\label{app:subtlety}

Imagine that a group has two generators $S_1$ and $S_2$ which do not commute.
Then ${\cal S}=\{S_1, S_2\}$ is a non-Abelian subset of
$\mathbb{U}(N)$.
But the relevant group of transformations of the scalar potential,
the so-called projective group, could be Abelian.
In fact, as pointed out in detail in
\cite{Ivanov:2011ae,Ivanov:2012ry,Ivanov:2012fp},
the group of physically distinct unitary reparametrization transformations
is not $\mathbb{SU}(N)$, but rather the projective special unitary group
$\mathbb{PSU}(N) \simeq \mathbb{SU}(N)/\mathbb{Z}_N
\simeq \mathbb{U}(N)/\mathbb{U}(1)$.
As an example, take a 3HDM with the symmetry
$\Delta(27)$ generated by
\begin{equation}
a=\left(
\begin{array}{ccc}
1 & 0 & 0\\
0 & \gamma & 0\\
0 & 0 & \gamma^2
\end{array}
\right),
\ \ \ \ 
b=\left(
\begin{array}{ccc}
0 & 1 & 0\\
0 & 0 & 1\\
1 & 0 & 0
\end{array}
\right),
\label{ab}
\end{equation}
with $\gamma=\exp{(2 i \pi/3)}$.
This is a non-Abelian subgroup of  $\mathbb{SU}(3)$.
Nevertheless,
the relevant structure to describe the invariances of the Higgs potential
is not $\mathbb{SU}(3)$ but rather
$\mathbb{PSU}(3) \simeq \mathbb{SU}(3)/\mathbb{Z}_3$.
And, since
\begin{equation}
a\, b\, a^{-1}\, b^{-1}
=
\omega^2
\left(
\begin{array}{ccc}
1 & 0 & 0\\
0 & 1 & 0\\
0 & 0 & 1
\end{array}
\right),
\end{equation}
the relevant group is rather
$\mathbb{Z}_3 \times \mathbb{Z}_3$,
which is Abelian \cite{Ivanov:2012ry}.
To make things slightly more complicated,
it turns out that creating a 3HDM invariant under
$\mathbb{Z}_3 \times \mathbb{Z}_3$ yields a scalar potential
invariant under the larger group
$(\mathbb{Z}_3 \times \mathbb{Z}_3) \rtimes \mathbb{Z}_2
\simeq \Sigma(54)/\mathbb{Z}_3$.
Further details can be found in ref.~\cite{Ivanov:2012ry}.

\section{\label{app:welldefined}3HDMs with a well-defined Symmetry basis}

Here we shall explicitly write the 3HDM scalar potential
for abelian symmetries that yield a well-defined Symmetry basis.
These are the $\mathbb{U}(1) \times \mathbb{U}(1)$,
$\mathbb{U}(1) \times \mathbb{Z}_2$, $\mathbb{Z}_3$ and $\mathbb{Z}_4$,
found in \cite{Ferreira:2008zy,Ivanov:2011ae},
and the $\mathbb{Z}_2 \times \mathbb{Z}_2$ put forward by
Weinberg \cite{Weinberg:1976hu}.
Such symmetries are realisable through the following representations,
\begin{gather}
    S_{\mathbb{U}(1) \times \mathbb{U}(1)}=\text{diag}(1,\,e^{i\alpha},\,e^{i\beta}), \label{eq:symmetry_generator_u1xu1}\\
    S_{\mathbb{U}(1) \times \mathbb{Z}_2} =\text{diag}(1,\,-1,\,e^{i\alpha}),
\label{eq:symmetry_generator_u1xz2}\\
    S_{\mathbb{Z}_2  \times \mathbb{Z}_2} =\left\{ \text{diag}(1,\,-1,\,1),\,\text{diag}(-1,\,1,\,1) \right\}, \label{eq:symmetry_generator_z2xz2}\\
S_{\mathbb{Z}_3}=\text{diag}(1,e^{i2\pi/3},e^{-i2\pi/3}),
\label{eq:symmetry_generator_z3}\\
S_{\mathbb{Z}_4}=\text{diag}(1,e^{i\pi},e^{-i\pi/2}).
\label{eq:symmetry_generator_z4}
\end{gather}
Here one can explicitly see that every doublet has a different
group charge, so that these symmetry groups have a well-defined
Symmetry basis.

By writing the general $\mathbb{U}(3)$ matrix in a suitable
diagonal basis, we obtain the generator of the
$\mathbb{U}(1) \times \mathbb{U}(1)$ symmetry
in \eqref{eq:symmetry_generator_u1xu1}.
Then the parameters that are invariant under
\eqref{eq:symmetry_generator_u1xu1}
will also be invariant under abelian symmetries whose generator
is written in a diagonal basis.
The $\mathbb{U}(1) \times \mathbb{U}(1)$ symmetric 3HDM can
be parameterised as \cite{Ivanov:2011ae},
\begin{align}
V_0 =&\, m_{11}^2\left(\phi_1^\dagger\phi_1\right)
	+ m_{22}^2\left(\phi_2^\dagger\phi_2\right)
	+ m_{33}^2\left(\phi_3^\dagger\phi_3\right)
\nonumber\\
& + \frac{\lambda_{11}}{2}\left(\phi_1^\dagger\phi_1\right)^2
	+ \frac{\lambda_{22}}{2}\left(\Phi_2^\dagger\Phi_2\right)^2
	+ \frac{\lambda_{33}}{2}\left(\Phi_3^\dagger\Phi_3\right)^2
\nonumber\\
&+ \lambda_{12}\left(\Phi_1^\dagger\Phi_1\right) \left(\Phi_2^\dagger\Phi_2\right)
	+ \lambda_{13}\left(\Phi_1^\dagger\Phi_1\right)\left(\Phi_3^\dagger\Phi_3\right)
\nonumber\\
&+ \lambda_{23}\left(\Phi_2^\dagger\Phi_2\right)\left(\Phi_3^\dagger\Phi_3\right)
	+ \lambda_{12}^{\prime}\left|\Phi_1^\dagger\Phi_2\right|^2
\nonumber\\
&+ \lambda_{13}^{\prime}\left|\Phi_1^\dagger\Phi_3\right|^2
	+ \lambda_{23}^{\prime}\left|\Phi_2^\dagger\Phi_3\right|^2.
\label{eq:V0_potential_3hdm}
\end{align}
The invariant potentials under each of the symmetry group
generators in \eqref{eq:symmetry_generator_u1xz2}-\eqref{eq:symmetry_generator_z4}
can be parameterised as
\cite{Ivanov:2011ae},
\begin{align}
&V_{\mathbb{U}(1) \times \mathbb{Z}_2} = V_0 +
	\frac{1}{2}\left[ \lambda_{12,12}\left(\Phi_1^\dagger\Phi_2\right)^2
	+ \text{h.c.} \right],
\label{eq:exact_potential_U1xz2}\\
&V_{\mathbb{Z}_2 \times \mathbb{Z}_2} = V_0
	+ \frac{1}{2}\left[ \lambda_{12,12}\left(\Phi_1^\dagger\Phi_2\right)^2
	+ \lambda_{13,13}\left(\Phi_1^\dagger\Phi_3\right)^2
	\right.
\nonumber\\
& \left. 
	\hspace{26mm} +\lambda_{23,23}\left(\Phi_2^\dagger\Phi_3\right)^2 + \text{h.c.} \right],
\label{eq:exact_potential_z2xz2}\\
&V_{\mathbb{Z}_3} = V_0 + \left[
	\lambda_{21,31}\left(\Phi_2^\dagger\Phi_1\right)\left(\Phi_3^\dagger\Phi_1\right)
	\right.
\nonumber\\
& \hspace{18mm} + \lambda_{12,32}\left(\Phi_1^\dagger\Phi_2\right)\left(\Phi_3^\dagger\Phi_2\right)
\nonumber\\
& \left.
	\hspace{18mm}
	+ \lambda_{13,23}\left(\Phi_1^\dagger\Phi_3\right)\left(\Phi_2^\dagger\Phi_3\right)
	+ \text{h.c.} \right],
\label{eq:exact_potential_z3} \\
&V_{\mathbb{Z}_4} = V_0
	+ \left[\lambda_{13,23}\left(\Phi_1^\dagger\Phi_3\right)\left(\Phi_2^\dagger\Phi_3\right)
	\right.
\nonumber\\
&\left.
	\hspace{18mm}
	+ \frac{1}{2}\lambda_{12,12}\left(\Phi_1^\dagger\Phi_2\right)^2
	+ \text{h.c.} \right].
\label{eq:exact_potential_z4}
\end{align} 
In practice, to make calculations with these symmetric potentials
it is convenient to use the tensorial parameterisation of the scalar
potential, with $Z^S_{ij,kl}=\lambda_{ij,kl}/2$.

In the notation of ref.~\cite{Ferreira:2008zy},
$\lambda_{11}=2 r_1$,
$\lambda_{22}=2 r_2$,
$\lambda_{33}=2 r_3$,
$\lambda_{12}=2 r_4$,
$\lambda_{13}=2 r_5$,
$\lambda_{23}=2 r_6$,
$\lambda_{12}^\prime =2 r_7$,
$\lambda_{13}^\prime =2 r_8$,
$\lambda_{23}^\prime =2 r_9$,
$\lambda_{12,12}=2 c_3$,
$\lambda_{13,13}=2 c_5$,
and
$\lambda_{23,23}=2 c_{17}$.


\begin{thebibliography}{99}
%
\bibitem{Arnison:1983rp}
  G.~Arnison {\it et al.} [UA1 Collaboration],
  ``Experimental Observation of Isolated Large Transverse
Energy Electrons with Associated Missing Energy at
s**(1/2) = 540-GeV,''
  Phys.\ Lett.\  {\bf 122B} (1983) 103.
  doi:10.1016/0370-2693(83)91177-2
%
\bibitem{Arnison:1983mk}
  G.~Arnison {\it et al.} [UA1 Collaboration],
  ``Experimental Observation of Lepton Pairs of Invariant
Mass Around 95-GeV/c**2 at the CERN SPS Collider,''
  Phys.\ Lett.\  {\bf 126B} (1983) 398.
  doi:10.1016/0370-2693(83)90188-0
%
\bibitem{Bagnaia:1983zx}
  P.~Bagnaia {\it et al.} [UA2 Collaboration],
  ``Evidence for Z0 $\rightarrow$ e+ e- at the CERN anti-p p Collider,''
  Phys.\ Lett.\  {\bf 129B} (1983) 130.
  doi:10.1016/0370-2693(83)90744-X
%
\bibitem{Banner:1983jy}
  M.~Banner {\it et al.} [UA2 Collaboration],
  ``Observation of Single Isolated Electrons of High
Transverse Momentum in Events with Missing Transverse
Energy at the CERN anti-p p Collider,''
  Phys.\ Lett.\  {\bf 122B} (1983) 476.
  doi:10.1016/0370-2693(83)91605-2
%
\bibitem{ALEPH:2005ab}
S.~Schael {\it et al.}
[ALEPH and DELPHI and L3 and OPAL and SLD Collaborations and LEP Electroweak Working Group and
SLD Electroweak Group and SLD Heavy Flavour Group],
``Precision electroweak measurements on the $Z$ resonance,''
Phys.\ Rept.\  {\bf 427} (2006) 257, doi:10.1016/j.physrep.2005.12.006
[hep-ex/0509008].
%
\bibitem{Aad:2012tfa}
G.~Aad {\it et al.} [ATLAS Collaboration],
``Observation of a new particle in the search for the
Standard Model Higgs boson with the ATLAS detector at the LHC,''
Phys.\ Lett.\ B {\bf 716} (2012) 1,
doi:10.1016/j.physletb.2012.08.020
[arXiv:1207.7214 [hep-ex]].
%
\bibitem{Chatrchyan:2012xdj}
S.~Chatrchyan {\it et al.} [CMS Collaboration],
``Observation of a New Boson at a Mass of 125 GeV with the CMS Experiment at the LHC,''
Phys.\ Lett.\ B {\bf 716} (2012) 30,
doi:10.1016/j.physletb.2012.08.021
[arXiv:1207.7235 [hep-ex]].
%
\bibitem{Khachatryan:2016vau}
G.~Aad {\it et al.} [ATLAS and CMS Collaborations],
``Measurements of the Higgs boson production and decay rates
and constraints on its couplings from a combined ATLAS and CMS analysis of
the LHC pp collision data at $ \sqrt{s}=7 $ and 8 TeV,''
JHEP {\bf 1608} (2016) 045,
doi:10.1007/JHEP08(2016)045
[arXiv:1606.02266 [hep-ex]].
%
\bibitem{Glashow:1976nt}
S.~L.~Glashow and S.~Weinberg,
``Natural Conservation Laws for Neutral Currents,''
Phys.\ Rev.\ D {\bf 15} (1977) 1958,
doi:10.1103/PhysRevD.15.1958
%
\bibitem{Gunion:2002zf}
J.~F.~Gunion and H.~E.~Haber,
``The CP conserving two Higgs doublet model: The Approach to the decoupling limit,''
Phys.\ Rev.\ D {\bf 67} (2003) 075019,
doi:10.1103/PhysRevD.67.075019
[hep-ph/0207010].
%
\bibitem{Nebot:2019lzf}
M.~Nebot,
``Non-Decoupling in Two Higgs Doublets Models, Spontaneous CP Violation
and $\mathbb{Z}_2$ symmetry,''
arXiv:1911.02266 [hep-ph].
%
\bibitem{Nierste:2019fbx}
U.~Nierste, M.~Tabet and R.~Ziegler,
``Cornering Spontaneous CP Violation with Charged-Higgs Searches,''
arXiv:1912.11501 [hep-ph].
%
\bibitem{Bhattacharyya:2014oka}
G.~Bhattacharyya and D.~Das,
``Nondecoupling of charged scalars in Higgs decay to
two photons and symmetries of the scalar potential,''
Phys.\ Rev.\ D {\bf 91} (2015) 015005,
doi:10.1103/PhysRevD.91.015005
[arXiv:1408.6133 [hep-ph]].
%
\bibitem{Botella:1994cs}
  F.~J.~Botella and J.~P.~Silva,
  ``Jarlskog - like invariants for theories with scalars and fermions,''
  Phys.\ Rev.\ D {\bf 51} (1995) 3870,
  doi:10.1103/PhysRevD.51.3870
  [hep-ph/9411288].
\bibitem{Barroso:2006pa}
  A.~Barroso, P.~M.~Ferreira, R.~Santos and J.~P.~Silva,
  ``Stability of the normal vacuum in multi-Higgs-doublet models,''
  Phys.\ Rev.\ D {\bf 74} (2006) 085016,
  doi:10.1103/PhysRevD.74.085016
  [hep-ph/0608282].
\bibitem{Bento:2017eti}
  M.~P.~Bento, H.~E.~Haber, J.~C.~Rom\~{a}o and J.~P.~Silva,
  ``Multi-Higgs doublet models: physical parametrization, sum rules and unitarity bounds,''
  JHEP {\bf 1711} (2017) 095,
  doi:10.1007/JHEP11(2017)095
  [arXiv:1708.09408 [hep-ph]].
%
\bibitem{Ferreira:2008zy}
  P.~M.~Ferreira and J.~P.~Silva,
  ``Discrete and continuous symmetries in multi-Higgs-doublet models,''
  Phys.\ Rev.\ D {\bf 78} (2008) 116007,
  doi:10.1103/PhysRevD.78.116007
  [arXiv:0809.2788 [hep-ph]].
%
\bibitem{Nishi:2007nh}
  C.~C.~Nishi,
  ``The Structure of potentials with N Higgs doublets,''
  Phys.\ Rev.\ D {\bf 76} (2007) 055013
  doi:10.1103/PhysRevD.76.055013
  [arXiv:0706.2685 [hep-ph]].
%
\bibitem{Belusca-Maito:2016dqe}
  H.~B\'{e}lusca-Ma\"{i}to, A.~Falkowski, D.~Fontes,
J.~C.~Rom\~{a}o and J.~P.~Silva,
  ``Higgs EFT for 2HDM and beyond,''
  Eur.\ Phys.\ J.\ C {\bf 77} (2017) no.3,  176
  doi:10.1140/epjc/s10052-017-4745-5
  [arXiv:1611.01112 [hep-ph]].
%
\bibitem{Bernon:2015qea}
  J.~Bernon, J.~F.~Gunion, H.~E.~Haber, Y.~Jiang and S.~Kraml,
  ``Scrutinizing the alignment limit in two-Higgs-doublet models: m$_h$=125  GeV,''
  Phys.\ Rev.\ D {\bf 92} (2015) no.7,  075004
  doi:10.1103/PhysRevD.92.075004
  [arXiv:1507.00933 [hep-ph]].
%
\bibitem{Ivanov:2011ae}
  I.~P.~Ivanov, V.~Keus and E.~Vdovin,
  ``Abelian symmetries in multi-Higgs-doublet models,''
  J.\ Phys.\ A {\bf 45} (2012) 215201
  doi:10.1088/1751-8113/45/21/215201
  [arXiv:1112.1660 [math-ph]].
%
\bibitem{Weinberg:1976hu}
  S.~Weinberg,
  ``Gauge Theory of CP Violation,''
  Phys.\ Rev.\ Lett.\  {\bf 37} (1976) 657.
  doi:10.1103/PhysRevLett.37.657
%
\bibitem{Branco:1979pv}
  G.~C.~Branco,
  ``Spontaneous CP Violation in Theories with More Than Four Quarks,''
  Phys.\ Rev.\ Lett.\  {\bf 44} (1980) 504.
  doi:10.1103/PhysRevLett.44.504
%
\bibitem{Lavoura:1994fv}
  L.~Lavoura and J.~P.~Silva,
  ``Fundamental CP violating quantities in a SU(2) x U(1) model
  with many Higgs doublets,''
  Phys.\ Rev.\ D {\bf 50} (1994) 4619
  doi:10.1103/PhysRevD.50.4619
  [hep-ph/9404276].
%
\bibitem{Branco:1999fs}
  G.~C.~Branco, L.~Lavoura and J.~P.~Silva,
  ``CP Violation,''
  Int.\ Ser.\ Monogr.\ Phys.\  {\bf 103} (1999) 1.
%
\bibitem{Davidson:2005cw}
S.~Davidson and H.~E.~Haber,
``Basis-independent methods for the two-Higgs-doublet model,''
Phys. Rev. D \textbf{72} (2005), 035004
doi:10.1103/PhysRevD.72.099902
[arXiv:hep-ph/0504050 [hep-ph]].
%
\bibitem{Gunion:2005ja}
J.~F.~Gunion and H.~E.~Haber,
``Conditions for CP-violation in the general two-Higgs-doublet model,''
Phys. Rev. D \textbf{72} (2005), 095002
doi:10.1103/PhysRevD.72.095002
[arXiv:hep-ph/0506227 [hep-ph]].
%
\bibitem{Haber:2006ue}
H.~E.~Haber and D.~O'Neil,
Phys. Rev. D \textbf{74} (2006), 015018
doi:10.1103/PhysRevD.74.015018
[arXiv:hep-ph/0602242 [hep-ph]].
%
\bibitem{Haber:2010bw}
H.~E.~Haber and D.~O'Neil,
Phys. Rev. D \textbf{83} (2011), 055017
doi:10.1103/PhysRevD.83.055017
[arXiv:1011.6188 [hep-ph]].
%
\bibitem{Ivanov:2015mwl}
  I.~P.~Ivanov and J.~P.~Silva,
  ``$CP$-conserving multi-Higgs model with irremovable
  complex coefficients,''
  Phys.\ Rev.\ D {\bf 93} (2016) no.9,  095014
  doi:10.1103/PhysRevD.93.095014
  [arXiv:1512.09276 [hep-ph]].
%
\bibitem{Branco:1985aq}
  G.~C.~Branco and M.~N.~Rebelo,
  ``The Higgs Mass in a Model With Two Scalar Doublets and Spontaneous {CP} Violation,''
  Phys.\ Lett.\  {\bf 160B} (1985) 117
  doi:10.1016/0370-2693(85)91476-5.
%
\bibitem{2appear}
H. E. Haber and J. P. Silva (to appear).
%
\bibitem{Ivanov:2012ry}
  I.~P.~Ivanov and E.~Vdovin,
  ``Discrete symmetries in the three-Higgs-doublet model,''
  Phys.\ Rev.\ D {\bf 86} (2012) 095030
  doi:10.1103/PhysRevD.86.095030
  [arXiv:1206.7108 [hep-ph]].
%
\bibitem{Ivanov:2012fp}
  I.~P.~Ivanov and E.~Vdovin,
  ``Classification of finite reparametrization symmetry groups in the three-Higgs-doublet model,''
  Eur.\ Phys.\ J.\ C {\bf 73} (2013) no.2,  2309
  doi:10.1140/epjc/s10052-013-2309-x
  [arXiv:1210.6553 [hep-ph]].
%
\bibitem{Faro:2019vcd}
F.~S.~Faro and I.~P.~Ivanov,
``Boundedness from below in the $U(1)\times U(1)$ three-Higgs-doublet model,''
Phys. Rev. D \textbf{100} (2019) no.3, 035038
doi:10.1103/PhysRevD.100.035038
[arXiv:1907.01963 [hep-ph]].
%
\bibitem{Ivanov:2007de}
I.~P.~Ivanov,
``Minkowski space structure of the Higgs potential in 2HDM. II.
Minima, symmetries, and topology,''
Phys. Rev. D \textbf{77} (2008), 015017
doi:10.1103/PhysRevD.77.015017
[arXiv:0710.3490 [hep-ph]].
%
\bibitem{Ferreira:2009wh}
P.~Ferreira, H.~E.~Haber and J.~P.~Silva,
``Generalized CP symmetries and special regions of parameter
space in the two-Higgs-doublet model,''
Phys. Rev. D \textbf{79} (2009), 116004
doi:10.1103/PhysRevD.79.116004
[arXiv:0902.1537 [hep-ph]].
%
%
\bibitem{Maniatis:2006fs}
M.~Maniatis, A.~von Manteuffel, O.~Nachtmann and F.~Nagel,
Eur. Phys. J. C \textbf{48} (2006), 805-823
doi:10.1140/epjc/s10052-006-0016-6
[arXiv:hep-ph/0605184 [hep-ph]].
%
\end{thebibliography}
\end{document}